\pdfoutput=1
\documentclass{vgtc} 
\ifpdf
  \pdfoutput=1\relax                   
  \usepackage{graphicx}                
  \DeclareGraphicsExtensions{.pdf,.png,.jpg,.jpeg} 
\else
  \usepackage{graphicx}                
  \DeclareGraphicsExtensions{.eps}     
\fi%

\graphicspath{{figures/}{pictures/}{images/}{./}} 
\usepackage[monochrome]{xcolor}
\usepackage{microtype}                 
\PassOptionsToPackage{warn}{textcomp}  
\usepackage{textcomp}                  
\usepackage{mathptmx}                  
\usepackage{times}                     
\usepackage{cite}                      
\usepackage{tabu}                      
\usepackage{booktabs} 
\usepackage{algorithmic}
\usepackage{graphicx}
\usepackage{multirow}
\usepackage{dsfont}
\usepackage{amsmath}
\usepackage{breqn}
\usepackage[skip= 0pt]{caption}
\usepackage{nomencl}
\usepackage{etoolbox}
\usepackage{amssymb}
\usepackage[utf8]{inputenc}
\usepackage[english]{babel}
\usepackage{mathtools}
\usepackage[export]{adjustbox}
\usepackage{float}
\usepackage{caption}
\usepackage{subcaption}
\usepackage{lipsum}
\usepackage{multicol}
\usepackage{lettrine}
\usepackage{nomencl}
\usepackage{babel}
\usepackage{adjustbox}
\usepackage{collectbox}
\usepackage{lipsum}
\usepackage{booktabs}
\usepackage{comment}
\usepackage{epstopdf}
\usepackage[vlined,linesnumbered,ruled]{algorithm2e}
\SetKwInOut{Parameter}{Parameter}
\SetKwInOut{Output}{Output}

\onlineid{1792}

\vgtccategory{Research}
\vgtcpapertype{please specify}

\title{LiteVR: Interpretable and Lightweight Cybersickness Detection using Explainable AI}

\author{Ripan Kumar Kundu \thanks{e-mail: rkcgc@umsystem.edu}\\
        \scriptsize University of Missouri %
\and Rifatul Islam \thanks{e-mail: rislam11@kennesaw.edu}\\ %
     \scriptsize Kennesaw State University %
\and John Quarles \thanks{e-mail: john.quarles@utsa.edu}\\ %
     \scriptsize University of Texas at San Antonio
\and Khaza Anuarul Hoque \thanks{e-mail: hoquek@umsystem.edu}\\
     \scriptsize University of Missouri}

\abstract{

Cybersickness is a common ailment associated with virtual reality (VR) user experiences. Several automated methods exist based on machine learning (ML) and deep learning (DL) to detect cybersickness. However, most of these cybersickness detection methods are perceived as computationally intensive and black-box methods. \textcolor{blue}{Thus, those techniques are neither trustworthy nor practical for deploying on standalone energy-constrained VR head-mounted devices (HMDs)}. In this work, we present an explainable artificial intelligence (XAI)-based framework \emph{LiteVR} for cybersickness detection, explaining the model's outcome, reducing the feature dimensions, and overall computational costs. First, we develop three cybersickness DL models based on long-term short-term memory (LSTM), gated recurrent unit (GRU), and multilayer perceptron (MLP). Then, we employed a post-hoc explanation, such as  SHapley Additive Explanations (SHAP), to explain the results and extract the most dominant features of cybersickness. Finally, we retrain the DL models with the reduced number of features. Our results show that eye-tracking features are the most dominant for cybersickness detection. Furthermore, based on the XAI-based feature ranking and dimensionality reduction, we significantly reduce the model’s size by up to 4.3$\times$, training time by up to 5.6$\times$, and its inference time by up to 3.8$\times$, with higher cybersickness detection accuracy and low regression error (i.e., on Fast Motion Scale (FMS)). Our proposed lite LSTM model obtained an accuracy of $94\%$ in classifying cybersickness and regressing (i.e., FMS 1--10) with a Root Mean Square Error (RMSE) of $0.30$, which outperforms the state-of-the-art. Our proposed \emph{LiteVR} framework can help researchers and practitioners analyze, detect, and deploy their DL-based cybersickness detection models in standalone VR HMDs.

} 


\CCScatlist{Human-centered computing---Human computer interaction (HCI)---Interaction paradigms---Virtual reality; 
Human-centered computing---Human computer interaction (HCI)---HCI design and evaluation methods.
}

\begin{document}


\firstsection{Introduction}

\maketitle
The applications of virtual reality (VR) are very diverse and are rapidly increasing in areas such as
health care~\cite{health1,health2}, education~\cite{alcaniz2022eye}, military training~\cite{yi2019dexcontroller}, surgical training~\cite{munawar2022virtual}, and disaster management/public safety~\cite{disas1,disas2}. However, VR users often experience virtual reality sickness, or cybersickness, which hinders their immersive experience. Thus cybersickness has emerged as an important obstacle ~\cite{nam2022eye} to the wider acceptability of VR. Cybersickness can be defined as a set of unpleasant symptoms such as eyestrain, headache, nausea, disorientation, or even vomiting~\cite{CS1}. One of the popular techniques to detect cybersickness is to use post-immersive subjective questionnaires such as Simulator Sickness Questionnaire (SSQ) and the VR Sickness Questionnaire (VRSQ) \cite{sevinc2020psychometric}. In contrast, the Fast Motion Sickness Scale (FMS) \cite{keshavarz2011validating} can assess cybersickness severity during immersion. Note, FMS relies on user feedback, i.e., needs human intervention during the VR immersion~\cite{melo2018presence}. To overcome these limitations, deep learning (DL) and machine learning (ML) have recently become popular for cybersickness detection~\cite{qu2022bio, islam2021cybersense,islam2020deep, oh2021machine,jin2018automatic, jeong2019cybersickness, islam2021cybersickness}. 

State-of-the-art DL models can detect cybersickness with good accuracy. For instance, Islam et al.~\cite{islam2021cybersickness} proposed a cybersickness severity detection using a deep fusion network with an accuracy of $87.7\%$ from users' eye tracking and head tracking data. Other researchers used  electroencephalography (EEG)~\cite{9233366, yokota2005motion, qu2022bio}, stereoscopic video~\cite{padmanaban2018towards, islam2021cybersickness} and bio-physiological signal~\cite{islam2020deep, qu2022bio} data for detecting cybersickness with higher accuracy. Despite the great prospect of DL models in detecting cybersickness, these methods have several limitations:

\begin{itemize}
    \item \textcolor{blue}{Most of the cybersickness DL-based detection models rely on black-box models; thus, they lack explainability. The explainability of the DL models can significantly improve the model's understanding and provide insight into why and how the DL model arrived at a specific decision. Identifying and understanding important features leading to cybersickness can help designers to develop more effective cybersickness detection models.}

    \item \textcolor{blue}{The multimodal data generated by HMD's integrated sensors as well as external physiological sensors (EEG, GSR, HR) results in a complex dataset and power-hungry DL models for cybersickness detection. As a result, deploying these models on standalone energy-constrained HMDs (e.g., Meta Quest Pro) is often challenging and not feasible. By identifying the dominant features, it can be useful to design and develop lightweight DL models that can improve computational costs, inference time, and model complexity. 
    }
\end{itemize}

\textcolor{blue}{To demonstrate the need for the proposed XAI approach, consider that a VR game developer may want to develop a cybersickness prediction model for their game based on alpha/beta players who are using a Meta Quest Pro HMD. If the developer was using a black box cybersickness prediction model without explainability in the game, they could not easily determine which of the model’s features (e.g., pupil diameter) contributed to cybersickness prediction. Hence, they would have use trial and error to reduce the model size such that it would minimize resource usage on the already resource constrained Meta Quest Pro.
}

Limited research has been conducted to address these research gaps. Recently, Kundu et al.~\cite{kundu2022truvr} used inherently interpretable ML models for cybersickness detection and explanation using bio-physiological\cite{islam2020automatic} and gameplay datasets\cite{porcino2021identifying}. However, this work considered binary classifiers, which can only detect the presence or absence of cybersickness. Such binary classification for cybersickness detection is less effective for realistic VR applications. Additionally, inherently interpretable models are typically dependent on the data properties and thus suffer from the dimensionality problem~\cite{schmitt2007limitations}. \textcolor{blue}{For instance, decision tree-based inherently interpretable models can suffer from overfitting problems because of over-complex trees that do not generalize the data well, which may eventually lead to their poor performance in cybersickness classification \cite{qu2022bio,8549968}}. 

Several other works use feature selection or dimensionality reduction techniques, such as principal component analysis (PCA)~\cite{kim2008application,satu2020towards,martin2020virtual,kim2018virtual,stone2017psychometric} to address the high-dimensionality issue in DL models for cybersickness detection. For instance, Mawalid et al. in~\cite{mawalid2018classification} used time-domain feature extraction methods to extract the EEG statistical features for classifying cybersickness. Similarly, Lin et al.~\cite{lin2013eeg} applied ML models with PCA to extract the cybersickness-related features to predict the cybersickness level. Furthermore, Kottaimalai et al. in~\cite{kottaimalai2013eeg} used PCA to reduce dimensions, complexity, and computational time for EEG signals to detect cybersickness. \textcolor{blue}{Note, PCA attempts to cover as much variance as possible among the feature spaces in a dataset. If the number of principal components in a dataset is not selected carefully, it is possible to miss some information compared to the original set of features~\cite{alkinani2017patch}. Hence, applying PCA-based dimensionality reduction may result in losing important features essential for accurate cybersickness detection. }

To address the above-mentioned challenges, we propose a novel methodology, LiteVR--\textit{an XAI-based framework for cybersickness detection}, explanation, and \textcolor{blue}{feature size reduction (also known as dimensionality reduction)}. Specifically, we first develop three DL models for cybersickness detection based on long-term short-term memory (LSTM), gated recurrent unit (GRU), and multilayer perceptron (MLP). Then we employ post-hoc explanation techniques, namely SHapley Additive Explanations (SHAP)~\cite{lundberg2017unified}, to provide global and local explanations for analyzing, identifying, and ranking the dominant features causing cybersickness. \textcolor{blue}{The identified dominant features are then used to retrain the models, (i.e., to train them with a reduced number of features instead of using all the features)}. This makes a cybersickness detection model that is easy to use and has a much smaller number of trainable parameters. This makes training and inference faster while maintaining baseline accuracy. For instance, our results show that an LSTM model with all features classifies the cybersickness severity into 4 classes (none, low, medium, and high) with an accuracy of $92\%$ and regresses (FMS 1--10) the ongoing cybersickness with a Root Mean Square Error (RMSE) value of $0.36$. However, after the XAI-based feature reduction (i.e., with only 1/3 of the features of the baseline model) of the same LSTM model, we can classify the cybersickness severity and predict the ongoing cybersickness with an accuracy of $94\%$ and an RMSE of $0.30$, respectively. Furthermore, our XAI-based model reduction approach results in up to a 4.3$\times$ reduction in the model size (for LSTM) and up to a 5.5$\times$  and 3.8$\times$ improvement in training and inference time, respectively. Therefore, we believe that the proposed method can aid future researchers in comprehending, analyzing, and designing cybersickness detection models that are suitable for real-time cybersickness detection for standalone HMDs. \textcolor{blue}{To the best of our knowledge, this is the first work applying XAI for explaining and feature size reduction of DL-based cybersickness detection models}.

\begin{figure*}[t]
    \centering
    \includegraphics[width=0.75\textwidth]{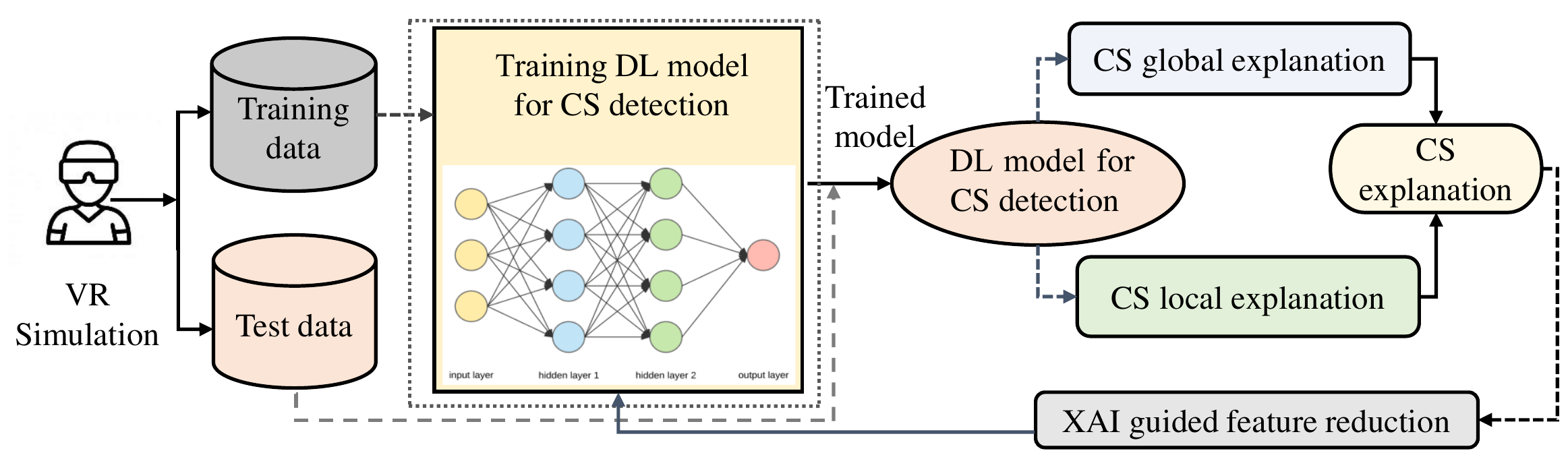}
    \caption{Overview of LiteVR for cybersickness (CS) detection, explanation, and dimensionality reduction.}
    \label{fig:method}
\end{figure*}

\section{Related Works}
\label{related}
The most popular theory to explain the reason behind cybersickness is the sensory conflict \cite{laviola2000discussion}. This theory states that cybersickness occurs when the visual sensory system perceives motion, but the vestibular system does not. However, other theories, such as poison theory and postural instability theory, even cybersecurity have also been identified as the causes of cybersickness\cite{laviola2000discussion, valluripally2021modeling, valluripally2022detection, gulhane2019security}. In addition, factors such as age, gender, and prior VR experience of users can also impact the degree of cybersickness \cite{garrido2022focusing, howard2021meta, 9779506}. The state-of-the-art works in cybersickness detection can be broadly divided into two categories that use either subjective and objective measurements. Recently, automated cybersickness detection using advanced ML and DL methods is getting more popular as it does not require any human interventions. In this section, we highlight the state-of-the-art works for cybersickness detection. 

To measure cybersickness, researchers have proposed several subjective measurements such as the Simulator Sickness Questionnaire (SSQ) \cite{chattha2020motion, zielasko2018dynamic, 9757545, 9760161, dong2022vr, sevinc2020psychometric, chan2022predicting, 9419347, oh2022cybersickness}, the FMS \cite{keshavarz2011validating}, and the Motion Sickness Susceptibility Questionnaire (MSSQ) \cite{kennedy1993simulator}. In contrast, several researchers have also proposed objective measurements (i.e., physiological signals) for cybersickness \cite{porcino2021identifying, islam2020automatic,islam2020deep} detection. Previous research has shown that objective measurements (e.g., heart rate, galvanic skin response, and electroencephalogram (EEG) signals) vary significantly when cybersickness occurs \cite{dennison2016use, islam2020automatic, lin201865, recenti2021toward, lin201865, reyero2022heart}. For example, they found that HR and EEG delta waves correlate positively with cybersickness, whereas EEG beta waves correlate negatively \cite{lin201865}. On the contrary, another study reported that GSR has a stronger positive correlation with cybersickness than other objective measurements that can detect cybersickness \cite{tian2022review, islam2020automatic}.

To automatically detect cybersickness from a variety of objective measurements (physiological signals) data, subjective measurements (FMSQ, SSQ) data and integrated sensors measurements (eye-tracking, head-tracking, motion-flow, etc.) data in HMD, numerous ML and DL-based approaches have recently been proposed \cite{qu2022bio, reddy2022estimating, reyero2022heart, islam2020automatic, islam2020deep, panahi2022simplified,keshavarz2022detecting,oh2021machine,ved2021detecting,wang2021using,islam2021cybersickness, yang2022machine, jeong2022leveraging, hadadi2022prediction}. For instance, in \cite{oh2021machine}, a machine–deep–ensemble learning method is applied to classify the cybersickness. In contrast, Porcino et al. \cite{porcino2021identifying} used a symbolic ML-based approach to identify the levels of cybersickness. In recent years DL has gained more attention from the cybersickness researchers. For instance, Qu et al. \cite{qu2022bio} used an LSTM-based attention network for detecting cybersickness using bio-physiological signals. On the other hand, Jin et al. \cite{jin2018automatic} utilized three DL/ML algorithms: Convolutional neural network (CNN), LSTM, and support vector machine (SVM) for cybersickness detection. Similarly, Padmanaban et al. \cite{padmanaban2018towards} used depth and optical flow features from the VR video data to predict cybersickness. In contrast, the authors in \cite{lee2019motion}  used a 3D-CNN and a multi-modal deep fusion approach with optical-flow, disparity, and saliency features and reported a better cybersickness detection accuracy compared to Padmanaban et al. in \cite{padmanaban2018towards}. Consequently, in \cite{islam2020automatic}, the authors used an LSTM model to classify the cybersickness severity from users' physiological signals (e.g., HR, GSR, etc.). In addition, a deep fusion approach was presented in \cite{islam2021cybersickness} for classifying cybersickness severity to forecast cybersickness from integrated sensors in HMD data. \textcolor{blue}{On the other hand, Jeong et al. \cite{jeong2022leveraging} used attention-based DL models for detecting cybersickness from multimodal sensor data.} Even though DL models have shown tremendous success in cybersickness detection, there is a significant research gap in applying XAI to explain cybersickness models. Indeed, understanding the feature importance in the cybersickness models and explaining why samples are being correctly vs. incorrectly labeled as cybersickness is an important step towards applying the proper mitigation technique. Furthermore, applying XAI in cybersickness models can significantly improve the model's \emph{trustworthiness} and provide insight into why and how the DL model arrived at a specific decision. In our context, the term trustworthy refers to using XAI for cybersickness explanation through a set of mechanisms, such as global and local explanations and explainable layers, to make the model transparent, understandable, and therefore, trusted by users~\cite{toreini2020relationship}. Very recently, Kundu et al.~\cite{kundu2022truvr} used inherently interpretable ML-based models for explaining cybersickness. They used explainable boosting Machine (EBM), decision tree (DT), and logistic regression (LR) to detect and explain the cybersickness from a user's bio-physiological and subjective measurement signals. But unfortunately, their proposed approach cannot explain DL-based cybersickness detection models. In contrast, we use a post-hoc explanation method based on SHAP to explain DL cybersickness detection models in our work. Another limitation in \cite{kundu2022truvr} is that their ML models are limited to binary classification (cybersickness vs. no cybersickness). In contrast, our work considers a multi-class classification problem (none, low, medium, and high) for cybersickness detection.

There exist several works in dimensionality reduction of cybersickness models. Several researchers used PCA-based methods to extract important features and reduce the model size~\cite{9417683, satu2020towards,stone2017psychometric, martin2020virtual,kim2018virtual,kim2008application} for this purpose. For instance, authors in~\cite{lin2013eeg}  used PCA to extract the cybersickness-related features from EEG signals. Then they used 3 ML/DL models: linear regression (LR), SVM, and self-organizing neural fuzzy inference network (SOFIN) to predict the user's level of cybersickness. 
Similarly, Kottaimalai et al. in~\cite{kottaimalai2013eeg} also used PCA to find the patterns from the EEG signals and neural networks (NN) to classify the cognitive tasks using the Colorado University EEG signal dataset. In contrast, Singla et al. in~\cite{9417683} used PCA to reduce the set of questions from the SSQ simulation. On the other hand, Mawalid et al. in~\cite{mawalid2018classification} used time domain feature extraction based on the statistical features (e.g., mean, variation, standard deviation, number of peaks) and power percentage band to understand the cybersickness features and then applied K-Nearest Neighbor and Naive Bayes classifiers to classify cybersickness. However, PCA-based dimension reduction is not always trustworthy. This is because PCA maps high-dimensional data to low-dimensional space through projections, which often cause the loss of information from the original data\cite{jia2022feature}. In contrast to that, we use a \emph{trustworthy}, i.e., XAI-based approach for dimensionality reduction of cybersickness detection models. 

\section{Proposed LiteVR framework}
An overview of the proposed LiteVR framework for VR cybersickness, detection, explanation, and model reduction is shown in \textcolor{blue}{Figure~\ref{fig:method}}. First, the training data \cite{islam2021cybersickness} is used to train the DL models (i.e., LSTM, GRU, and MLP) for both classification and regression tasks \footnote{Dataset \cite{islam2021cybersickness}: https://tinyurl.com/2p92p45h}. Next, the trained DL models detect cybersickness on the test dataset by classifying them into four classes-- none, low, medium, and high. The trained DL models can also forecast the next value of the ongoing cybersickness using the FMS score range of $0-10$. Next, we use the state-of-the-art tool SHAP to explain the cybersickness features, which can provide both local and global explanations. Global explanation identifies features crucial for the overall prediction, and local explanation identifies features dominating an individual sample prediction. After explaining the cybersickness features, we use the global explanation to identify and rank the dominating features. Finally, in the model reduction phase, we retrain the DL models with only the most dominating features (in our case, only the top 30\%) to reduce the cybersickness models. This results in a significantly smaller model with fewer trainable parameters, hence a faster training and inference time. 

\subsection{Cybersickness classification}
\label{los_dl}
The network architecture of the LSTM, GRU, and MLP-based cybersickness classification model is shown in Tables \ref{LSTM_without} , \ref{gru_without}, and  \ref{mlp_without}, respectively. The LSTM and GRU models have 6 layers, whereas the MLP model has 8 layers. We chose LSTM and GRU models as they can capture datasets' temporal and spatial features. For both the LSTM and GRU models, we set a timestep of 60 and have 43 features. The first LSTM layer had a recurrent dropout of $20\%$, followed by a same-dropout layer of $20\%$ to deal with overfitting. In the next LSTM layer, we reduce the dropout rate by up to $5\%$ because a high dropout rate can slow the model's convergence rate and often harm final performance \cite{wang2019jumpout}. On the other hand, for the GRU model, all of the GRU layers used a recurrent dropout of $30\%$ to reduce overfitting. For both the LSTM and GRU models, we used 'ReLu' as the activation function for the fifth dense layer. Finally, we used `softmax' as the activation function in the last dense layer, which contains four outputs for the four cybersickness classes. Since all the $8$ layers in MLP are dense, we do not have the timesteps like LSTM and GRU models. We use `ReLu' in each dense layer as an activation function with a recursive dropout of $25\%$. Like GRU and LSTM models, the last dense layer of MLP has the softmax activation function to identify the four cybersickness severity classes. We use categorical cross-entropy as the loss function in our DL classification models as shown in eqn.\ref{eqn1}. 

\begin{equation}\mathcal{L} (y,\displaystyle \hat{y})=-\sum_{i=1}^{p}\sum_{j=1}^{q}(y_{ij}\log(\hat{y}_{ij}))\label{eqn1} \end{equation}

where $\mathcal{L}$ denoted the loss function, $\hat{y}_{ij}$ is the predicted cybersickness severity class label, and $(y_{ij})$ is the actual cybersickness severity class label. Furthermore, $p$ is the number of cybersickness classes, and $q$ denotes the classifier's total number of training samples. 

\subsection{Cybersickness regression}
The cybersickness regression task can be defined as follows: Given a history of observed VR data (e.g., eye-tracking data, head-tracking data, etc.) and the FMS score at previous time steps $t-1$, predict the FMS score at the next time steps $t$. Let us denote  the cybersickness FMS score at time $t$ as  $CS_t$. For instance, if we predict the cybersickness FMS score at time $t = 10$ seconds, then $CS_{t} \rightarrow [S_{t-9}, S_{t-8}, S_{t-7}, S_{t-6} \ldots, S_{t}]$, where $S$ denotes the user’s cognitive state. We use the same three DL models that we used for classification for the cybersickness regression task as well except the loss function (RMSE for regression), i.e., to forecast the cybersickness FMS score in the range of $0$ to $10$.
For the regression models, we use root mean squared error (RMSE) as the loss function as defined in Equation \ref{eqn2}. 

\begin{equation}
    RMSE= \sqrt{\frac{1}{\vert N\vert} \sum\limits_{y\in S_{}}\sum\limits_{t=1}^{N}(y_{t}-\hat{y}_{t})^{2}} \label{eqn2}
\end{equation}

where $y_{t}$ is the input, e.g., verbally reported FMS score from the user at time $t$, and $\hat{y}_{t}$ is the predicted FMS score by the model at that time. Consequently, $N$ and $S$ represent the total number of samples and time steps.

\SetInd{0.5em}{0.5em}
\begin{algorithm}[!h] 
\caption{SHAP-based cybersickness feature reduction}
\label{SHAP method}
\DontPrintSemicolon
\SetKwInOut{Input}{Inputs}
\SetKwInOut{Output}{Outputs}
\Input{Cybersickness DL Model $\theta$,\\
Training dataset $X_{Train}$, \\
Testing dataset $X_{Test}$}
\Output{$F_{rank}$ = $\{~\}$}
\begin{algorithmic}[1]
\FOR {\textbf{each} feature $i$ $\in$ $X_{train}$ }{
    \STATE $Ex_{val}$ = ModelExplainer($\theta$, $X_{Train}$) \\
  \STATE $F_{imp}$ =  SHAPval($Ex_{val}$, $X_{Test}$ )\\
   \STATE $F_{rank}$ = $F_{rank} \bigcup F_{imp} $ \\
}\ENDFOR
\STATE \textbf{return} $F_{rank}$
\end{algorithmic}
\end{algorithm}
\subsection{Explaining DL models with SHAP}
In this work, we use the SHAP-based post-hoc explanation technique for explaining the cybersickness outcomes. SHAP is a feature importance explanation approach that works by assigning a feature significance value to each prediction. It is based on the mathematical foundation of Shapley values from cooperative game theory~\cite{lundberg2017unified}. For a given set of input samples (e.g., eye-tracking data, head-tracking data, etc.) and DL models ( e.g., LSTM, GRU, and MLP), the goal of SHAP is to explain the prediction of input samples by calculating the contribution of each feature to the prediction. In other words, SHAP values explain how a given feature value increases or decreases a model’s prediction. For example, given a cybersickness prediction model that predicts whether a sample has cybersickness, the SHAP explanation allows us to know to what extent a feature drove a given prediction. 
Such explanations can be global or local. The overall feature importance ranking (global explanation) of cybersickness can be visualized in terms of bar graphs in SHAP. On the contrary, for the local explanation, each sample is randomly chosen from the test dataset, which contains all the features and determines which features increase or reduce the likelihood of cybersickness. We use the \emph{Deep SHAP} method \cite{shrikumar2016not} for computing SHAP values from the DL-based cybersickness models. After that, these SHAP values are used to calculate the feature's importance during cybersickness severity prediction.

\subsection{Feature reduction through SHAP-based explanation}

Algorithm \ref{SHAP method} shows the overall method of feature reduction using SHAP. The algorithm takes a cybersickness detection model, training and testing datasets as input, and returns the feature ranking. For each feature in the dataset, given the model $\theta$, the algorithm uses the \texttt{ModelExplainer()} function to obtain the expected value $Ex_{val}$ (Lines 1-2). The \texttt{ModelExplainer()} function approximates the conditional expectations of SHAP values using the $X_{Train}$ samples~\cite{lundberg2017unified}. Next, the algorithm calculates the feature importance score $F_{imp}$ for each feature using the \texttt{SHAPvalue()} function (Line 3), which is then appended to the set $F_{rank}$ (Line 4). The \texttt{SHAPvalue()} function calculates the $F_{imp}$ value through calculating the mean average value for each feature. Finally, when this process is complete for all the features, the algorithm returns the set $F_{rank}$ (Line 6). Based on the ranked features  $F_{rank}$, we sort (in descending order, i.e., the top feature means the most important feature) them based on their importance and use only a portion of the important/top features for retraining the model. For instance, in our case, we only considered the top $1/3$ of the features in the ranked list to retrain the cybersickness DL models. This results in a  lighter model with significantly fewer trainable parameters and, thus, a faster training and inference time.  


 

\section{Dataset \& Experimental Setup}
\label{experiment}
\textcolor{blue}{This section explains the experimental setup and data we used to validate our proposed LiteVR framework. We used Scikit-Learn~\cite{pedregosa2011scikit} and TensorFlow-2.4~\cite{sergeev2018horovod} for training  and evaluating our DL models. For explaining the DL models, we used the SHAP \cite{lundberg2017unified} library. The DL models were trained on an Intel Core i9 Processor and 32GB RAM option with NVIDIA GeForce RTX 3080 Ti GPU.}


\subsection{Hyper-Parameter}
\label{sec:hyperparameter}
We used the Adam optimizer to optimize our DL models with epochs of $300$ and a batch size of $256$. For training the DL models, we used the learning rate of $0.001$. To prevent the model from overfitting, we deployed an early-stopping strategy with a patience value of $30$ while training the DL models. We used a $10$ fold cross-validation technique to train and test the performance of the DL models similar to \cite{bengio2003no} in which the dataset is partitioned into $k$ groups (i.e., in our case $k$ = 10). Only one partition out of $k$ is utilized for testing the model, while the remaining partitions are used for training. The method is repeated $k$ times, each time picking a new test partition and the remaining $(k-1)$ partition as a training dataset to eliminate bias.

\subsection{Dataset}
We used the integrated sensor dataset to validate the effectiveness of the proposed DL models. The integrated sensor dataset \cite{islam2021cybersickness} contains the eye tracking, head tracking, and physiological signals for $30$ participants immersed in 5 different VR simulations: Beach City, Road Side, Roller Coaster, SeaVoyage, and Furniture Shop. Eye tracking, head tracking, and physiological data consist of different subcategories. For instance, in the eye tracking data, the subcategories are Pupil Diameter (left), Pupil Position (x, y, z), Gaze Direction (x, y, z), Convergence Distance, and $\%$ of Eye Openness, and for the head tracking data Quaternion Rotation of X, Y, Z, and W axis, respectively. Similarly, for the physiological signals, the subcategories are electrodermal activity (EDA) and HR measurements. This dataset has a total of $20104$ samples recorded with a maximum of $7$ minutes of VR simulation. In addition, as mentioned earlier, the dataset contains four different cybersickness severity classes: none, low, medium, and high, and the FMS score ranges from $0$ to $10$, which is used for regression analysis. We used the  $70\%$ samples from this dataset for training the DL models and their remaining $30\%$ samples for testing.


\subsection{Performance Metrics}
The standard quality metrics such as accuracy, recall, precision, and F1-score are used to evaluate the performance of DL models (e.g., LSTM, GRU, and GRU) for cybersickness severity classification \cite{goutte2005probabilistic}. Similarly, standard loss functions like mean squared error (MSE), root-mean-squared error (RMSE), mean absolute error (MAE), and $R^2$ score are used to evaluate the cybersickness regression model's performance \cite{chicco2021coefficient}. For example, if $y_{t}$ and $\hat{y}_{t}$ denote the actual and predicted cybersickness of a candidate at time $t$, respectively, then the MAE can be defined as:

\begin{equation*} 
    MAE= \frac{1}{\vert N\vert}\sum\limits_{y\in S_{}}\sum\limits_{t=1}^{N}(y_{t}-\hat{y}_{t})
\end{equation*}

where $N$ and $S$ represent the total number of samples and time steps of all trajectories. The smaller the MAE, the better the regression model. If ${\sum({y_{t}}-\hat{y_{t}})^2}$ and ${\sum(y_{t}-\bar{y})^2}$ represent the Sum Squared Regression (SSR) and the Total Sum of Squares (SST) then, the $R^2$ score can be written as: 
\begin{equation*} R^2=1-\frac{\sum({y_{t}}-\hat{y_{t}})^2}{\sum(y_{t}-\bar{y})^2}\end{equation*}

Consequently, a low $R^2$ value indicates that the regression model does not adequately capture the output variance.

\begin{table}[]
\centering
\caption{LSTM model architecture (with all features)}
\label{LSTM_without}
\resizebox{\columnwidth}{!}{%
\begin{tabular}{|ccc|ccc|}
\hline
\multicolumn{1}{|c|}{\textbf{Layer}} & \multicolumn{1}{c|}{\textbf{Type}} & \textbf{Output shape} & \multicolumn{1}{c|}{\textbf{\#Param}} & \multicolumn{1}{c|}{\textbf{Dropout}} & \textbf{Activation} \\ \hline
\multicolumn{1}{|c|}{1} & \multicolumn{1}{c|}{LSTM} & 128 & \multicolumn{1}{c|}{66560} & \multicolumn{1}{c|}{0.2} & - \\ \hline
\multicolumn{1}{|c|}{2} & \multicolumn{1}{c|}{Dropout} & 128 & \multicolumn{1}{c|}{0} & \multicolumn{1}{c|}{0.2} & - \\ \hline
\multicolumn{1}{|c|}{3} & \multicolumn{1}{c|}{LSTM} & 64 & \multicolumn{1}{c|}{49408} & \multicolumn{1}{c|}{0.15} &  -\\ \hline
\multicolumn{1}{|c|}{4} & \multicolumn{1}{c|}{Dropout} & 64 & \multicolumn{1}{c|}{0} & \multicolumn{1}{c|}{0.15} &  -\\ \hline
\multicolumn{1}{|c|}{5} & \multicolumn{1}{c|}{Dense} & 16 & \multicolumn{1}{c|}{3136} & \multicolumn{1}{c|}{-} & ReLU \\ \hline
\multicolumn{1}{|c|}{6} & \multicolumn{1}{c|}{Dense} & 4 & \multicolumn{1}{c|}{68} & \multicolumn{1}{c|}{-} & Softmax \\ \hline
\multicolumn{3}{|c|}{\textbf{Total no. of parameter:}} & \multicolumn{3}{l|}{\textbf{119, 172}} \\ \hline
\end{tabular}%
}
\end{table}

\begin{table}[t]
\centering
\caption{GRU model architecture (with all features)}
\label{gru_without}
\resizebox{\columnwidth}{!}{%
\begin{tabular}{|ccc|ccc|}
\hline
\multicolumn{1}{|c|}{\textbf{Layer}} & \multicolumn{1}{c|}{\textbf{Type}} & \textbf{Output shape} & \multicolumn{1}{c|}{\textbf{\#Param}} & \multicolumn{1}{c|}{\textbf{Dropout}} & \textbf{Activation} \\ \hline
\multicolumn{1}{|c|}{1} & \multicolumn{1}{c|}{GRU} & 32 & \multicolumn{1}{c|}{3360} & \multicolumn{1}{c|}{0.3} &  -\\ \hline
\multicolumn{1}{|c|}{2} & \multicolumn{1}{c|}{Dropout} & 32 & \multicolumn{1}{c|}{0} & \multicolumn{1}{c|}{0.3} &  -\\ \hline
\multicolumn{1}{|c|}{3} & \multicolumn{1}{c|}{GRU} & 64 & \multicolumn{1}{c|}{18816} & \multicolumn{1}{c|}{0.3} &  -\\ \hline
\multicolumn{1}{|c|}{4} & \multicolumn{1}{c|}{Dropout} & 64 & \multicolumn{1}{c|}{0} & \multicolumn{1}{c|}{0.3} &  -\\ \hline
\multicolumn{1}{|c|}{5} & \multicolumn{1}{c|}{Dense} & 16 & \multicolumn{1}{c|}{74496} & \multicolumn{1}{c|}{-} & ReLU \\ \hline
\multicolumn{1}{|c|}{6} & \multicolumn{1}{c|}{Dense} & 4 & \multicolumn{1}{c|}{516} & \multicolumn{1}{c|}{-} & Softmax \\ \hline
\multicolumn{3}{|c|}{\textbf{Total no. of param:}} & \multicolumn{3}{l|}{\textbf{97,188}} \\ \hline
\end{tabular}%
}
\end{table}
\begin{table}[t]
\centering
\caption{MLP model architecture (with all features)}
\label{mlp_without}
\resizebox{\columnwidth}{!}{%
\begin{tabular}{|ccc|ccc|}
\hline
\multicolumn{1}{|c|}{\textbf{Layer}} & \multicolumn{1}{c|}{\textbf{Type}} & \textbf{Output shape} & \multicolumn{1}{c|}{\textbf{\#Param}} & \multicolumn{1}{c|}{\textbf{Dropout}} & \textbf{Activation} \\ \hline
\multicolumn{1}{|c|}{1} & \multicolumn{1}{c|}{Dense} & 128 & \multicolumn{1}{c|}{5504} & \multicolumn{1}{c|}{0.25} & ReLU \\ \hline
\multicolumn{1}{|c|}{2} & \multicolumn{1}{c|}{Dropout} & 128 & \multicolumn{1}{c|}{0} & \multicolumn{1}{c|}{0.25} &  \\ \hline
\multicolumn{1}{|c|}{3} & \multicolumn{1}{c|}{Dense} & 64 & \multicolumn{1}{c|}{8256} & \multicolumn{1}{c|}{0.25} & ReLU \\ \hline
\multicolumn{1}{|c|}{4} & \multicolumn{1}{c|}{Dropout} & 64 & \multicolumn{1}{c|}{} & \multicolumn{1}{c|}{0.25} & - \\ \hline
\multicolumn{1}{|c|}{5} & \multicolumn{1}{c|}{Dense} & 32 & \multicolumn{1}{c|}{2080} & \multicolumn{1}{c|}{0.25} & ReLU \\ \hline
\multicolumn{1}{|c|}{6} & \multicolumn{1}{c|}{Dropout} & 32 & \multicolumn{1}{c|}{0} & \multicolumn{1}{c|}{0.25} & - \\ \hline
\multicolumn{1}{|c|}{7} & \multicolumn{1}{c|}{Dense} & 16 & \multicolumn{1}{c|}{1056} & \multicolumn{1}{c|}{-} & ReLU \\ \hline
\multicolumn{1}{|c|}{8} & \multicolumn{1}{c|}{Dense} & 4 & \multicolumn{1}{c|}{132} & \multicolumn{1}{c|}{-} & Softmax \\ \hline
\multicolumn{3}{|c|}{\textbf{Total no. of param:}} & \multicolumn{3}{l|}{\textbf{17,028}} \\ \hline
\end{tabular}%
}
\end{table}

\begin{table}[t]
\centering
\caption{Training and inference time of: non-reduced DL models (all features) vs. reduced DL models (reduced features)}
\label{train_time}
\resizebox{\columnwidth}{!}{%
\begin{tabular}{|c|cc|cc|}
\hline
\multirow{2}{*}{\textbf{ Models}} & \multicolumn{2}{c|}{\textbf{ Before XAI}} & \multicolumn{2}{c|}{\textbf{After XAI}} \\ \cline{2-5} 
 & \multicolumn{1}{c|}{\textbf{Training(s)}} & \textbf{Inference(s)} & \multicolumn{1}{c|}{\textbf{Training(s)}} & \textbf{Inference(s)} \\ \hline
\textbf{LSTM} & \multicolumn{1}{c|}{2076.21} & 4.05 & \multicolumn{1}{c|}{376.37} & 1.07 \\ \hline
\textbf{GRU} & \multicolumn{1}{c|}{3236.21} & 4.94 & \multicolumn{1}{c|}{774.29} & 1.32 \\ \hline
\textbf{MLP} & \multicolumn{1}{c|}{51.25} & 0.15 & \multicolumn{1}{c|}{21.55} & 0.09 \\ \hline
\end{tabular}%
}
\end{table}

\begin{table*}[ht]
\centering
\caption{Performance of $10$-Fold Cross Validation on Cybersickness Severity Classification (non-reduced DL models)}\label{acc}

\begin{tabular}{|c|c|cccc|cccc|cccc|}
\hline
\multirow{3}{*}{\textbf{Model}} & \multirow{3}{*}{\textbf{Accuracy\%}} & \multicolumn{4}{c|}{\textbf{Precision\%}} & \multicolumn{4}{c|}{\textbf{Recall \%}} & \multicolumn{4}{c|}{\textbf{F1-score\%}} \\ \cline{3-14} 
 &  & \multicolumn{1}{c|}{\textbf{None}} & \multicolumn{1}{c|}{\textbf{Low}} & \multicolumn{1}{c|}{\textbf{Medium}} & \textbf{High} & \multicolumn{1}{c|}{\textbf{None}} & \multicolumn{1}{c|}{\textbf{Low}} & \multicolumn{1}{c|}{\textbf{Medium}} & \textbf{High} & \multicolumn{1}{c|}{\textbf{None}} & \multicolumn{1}{c|}{\textbf{Low}} & \multicolumn{1}{c|}{\textbf{Medium}} & \textbf{High} \\ \hline
\textbf{LSTM} & 92 & \multicolumn{1}{c|}{96} & \multicolumn{1}{c|}{86} & \multicolumn{1}{c|}{87} & 90 & \multicolumn{1}{c|}{92} & \multicolumn{1}{c|}{95} & \multicolumn{1}{c|}{86} & 93 & \multicolumn{1}{c|}{95} & \multicolumn{1}{c|}{92} & \multicolumn{1}{c|}{83} & 90 \\ \hline
\textbf{GRU} & 90 & \multicolumn{1}{c|}{95} & \multicolumn{1}{c|}{89} & \multicolumn{1}{c|}{81} & 83 & \multicolumn{1}{c|}{94} & \multicolumn{1}{c|}{90} & \multicolumn{1}{c|}{79} & 87 & \multicolumn{1}{c|}{95} & \multicolumn{1}{c|}{90} & \multicolumn{1}{c|}{81} & 86 \\ \hline
\textbf{MLP} & 79 & \multicolumn{1}{c|}{83} & \multicolumn{1}{c|}{79} & \multicolumn{1}{c|}{80} & 75 & \multicolumn{1}{c|}{81} & \multicolumn{1}{c|}{85} & \multicolumn{1}{c|}{72} & 78 & \multicolumn{1}{c|}{82} & \multicolumn{1}{c|}{77} & \multicolumn{1}{c|}{70} & 75 \\ \hline
\end{tabular}%
\vspace{2mm}
\end{table*}

\begin{table}[ht!]
\centering
\caption{Cybersickness regression using non-reduced DL models (with all features)}
\begin{tabular}{|c|c|c|c|c|}
\hline
\textbf{Regression Models} & \textbf{MSE} & \textbf{RMSE} & \textbf{$R^2$} & \textbf{MAE} \\ \hline
LSTM & 0.24 & 0.36 & 90 & 0.21 \\ \hline
GRU & 0.29 & 0.41 & 86 & 0.24 \\ \hline
MLP & 0.42 & 0.57 & 73 & 0.40 \\ \hline
\end{tabular}%
\label{tab:reg-table}
\end{table}

\section{Results}
This section presents the results of cybersickness detection, explanation, and reduction.


\subsection{Cybersickness classification and regression performance with all features}
Before presenting the results of our LiteVR approach for cybersickness explanation and model reduction, in this section, we first present the details of the model development with their important statistics, cybersickness classification and regression accuracy. This will help us compare our approach's effectiveness in the following subsection of the paper. 

\subsubsection{Cybersickness detection model development}

Tables \ref{LSTM_without},  \ref{gru_without},  \ref{mlp_without} show the network architecture and number of trainable parameters for the LSTM, GRU, and MLP based cybersickness detection models. We observe that the ML, LSTM, and GRU model has total $17,028$, $119,172$, and $97,188$ trainable parameters. The MLP model has $7$ times and $5.7$ times less number of trainable parameters when compared to the LSTM and GRU models. These models' training and inference times are reported in Table \ref{train_time}. Training the MLP, LSTM, and GRU models (with all the features) requires $51.25$, $2076.21$, and $3236.21$ seconds, respectively. Indeed, training the MLP model is $232.33$ and $63.14$  times faster than the GRU and LSTM models. Similarly, the inference time required for the MLP, LSTM, and GRU models are $0.15$, $4.05$, and $4.94$ seconds, respectively. Indeed the inference on MLP is $27$ and $33$ times faster than the LSTM and GRU models, respectively.

\subsubsection{Cybersickness classification and regression with all features}\label{subsubsec:class}

Table \ref{acc} summarizes the accuracy, precision, recall, and F-1 scores of cybersickness severity classification using LSTM, GRU, and MLP models. The cybersickness classification using the LSTM and GRU model resulted in an accuracy of $92\%$ and $90\%$, whereas the accuracy of MLP is only $79\%$. Indeed the overall performance of the LSTM model is slightly better than the GRU model in terms of precision, recall, and F1-score for cybersickness severity classification. Furthermore, Table \ref{tab:reg-table} summarizes the MSE, RMSE, R\textsuperscript{2}, and MAE of cybersickness regression using LSTM, GRU and MLP models. Like cybersickness classification, regression using the LSTM outperforms the GRU and MLP models. Like classification, the MLP also performs poorly for regression task. 


\begin{figure*}[!t]
\vspace{-0.15in}
\centering      
\subfloat[\centering\label{fig:3a}]{\includegraphics[width=0.5\textwidth]{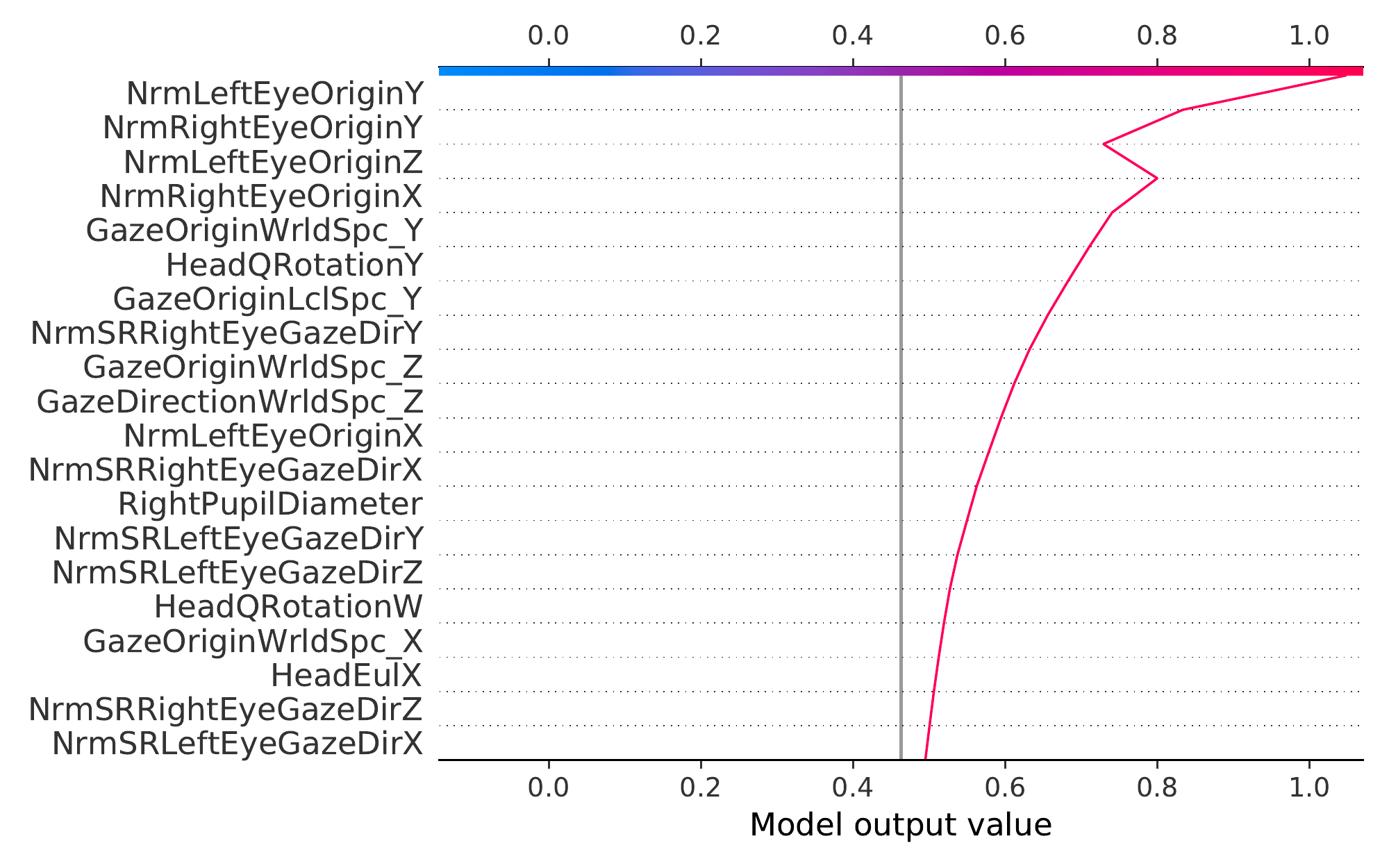}}\hfill
\subfloat[\centering\label{fig:3b}]{\includegraphics[width=0.5\textwidth]{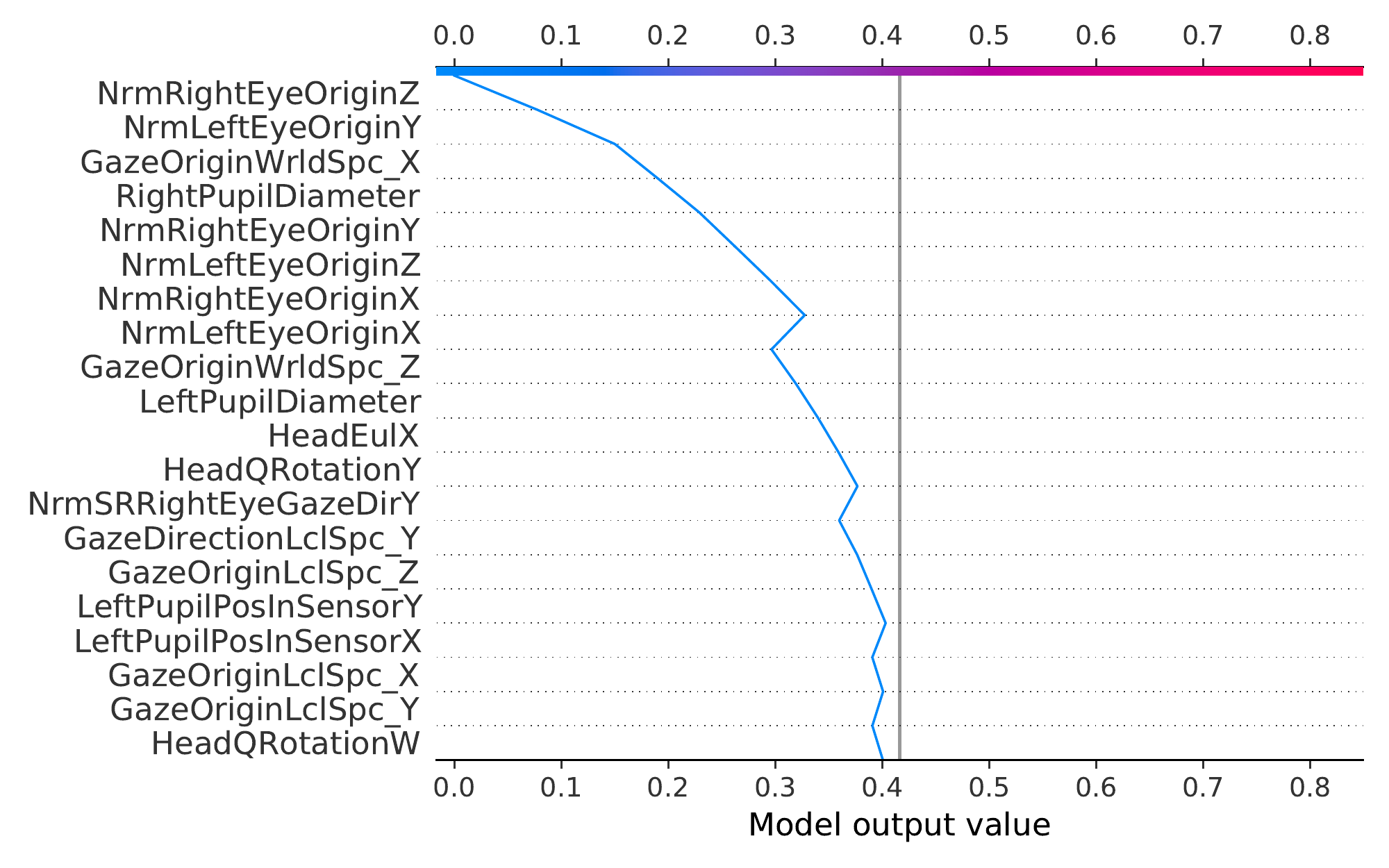}}
\vspace{5pt}
\caption{The local explanation of LSTM-based cybersickness classification for the (\textbf{a}) explanation for high cybersickness severity, (\textbf{b}) explanation for none cybersickness severity.} \label{fig:local_actual_LSTM}
\end{figure*}

\begin{figure*}[!t]
\vspace{-0.15in}
\centering      
\subfloat[\centering\label{fig:3a}]{\includegraphics[width=0.5\textwidth]{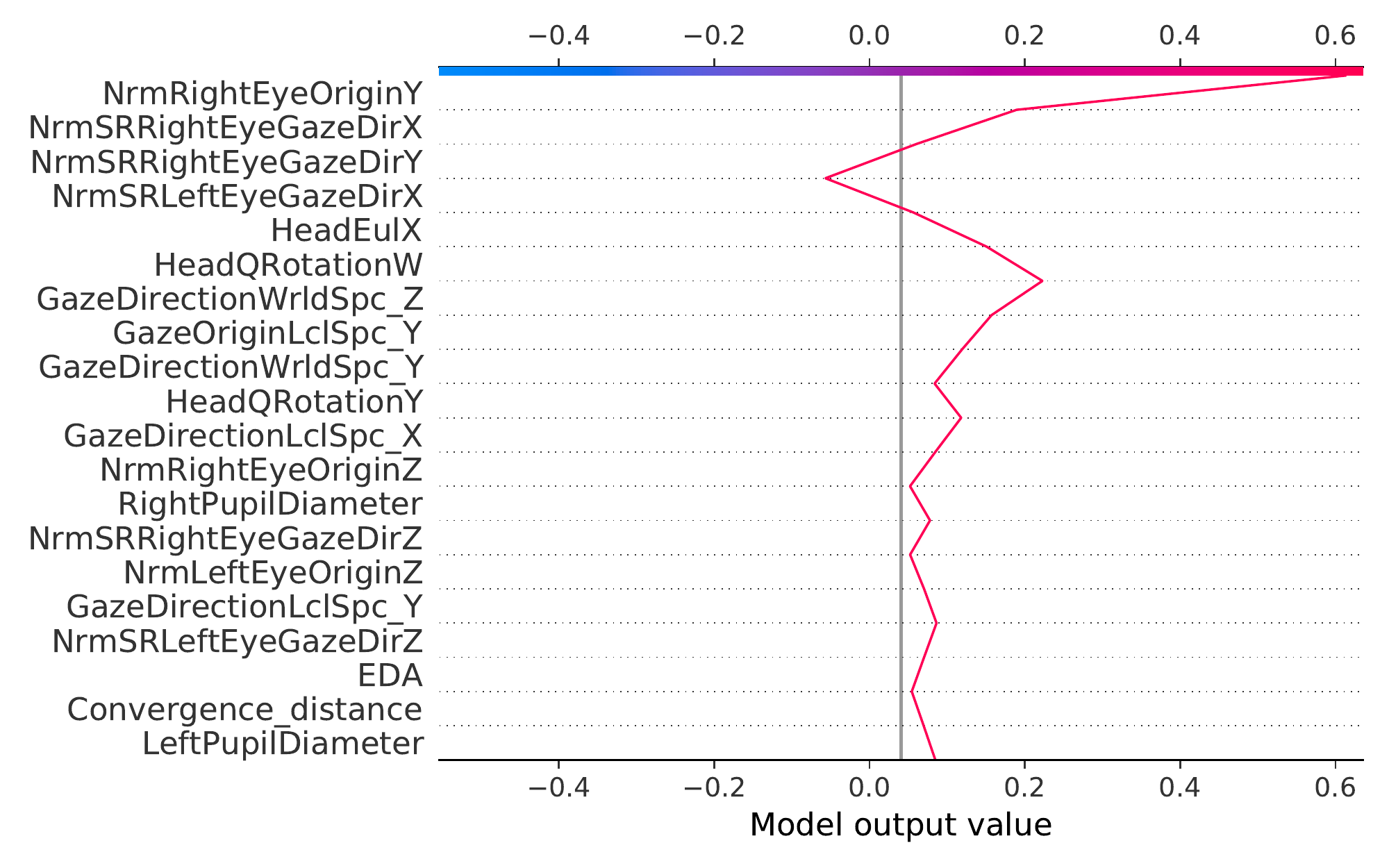}}\hfill
\subfloat[\centering\label{fig:3b}]{\includegraphics[width=0.5\textwidth]{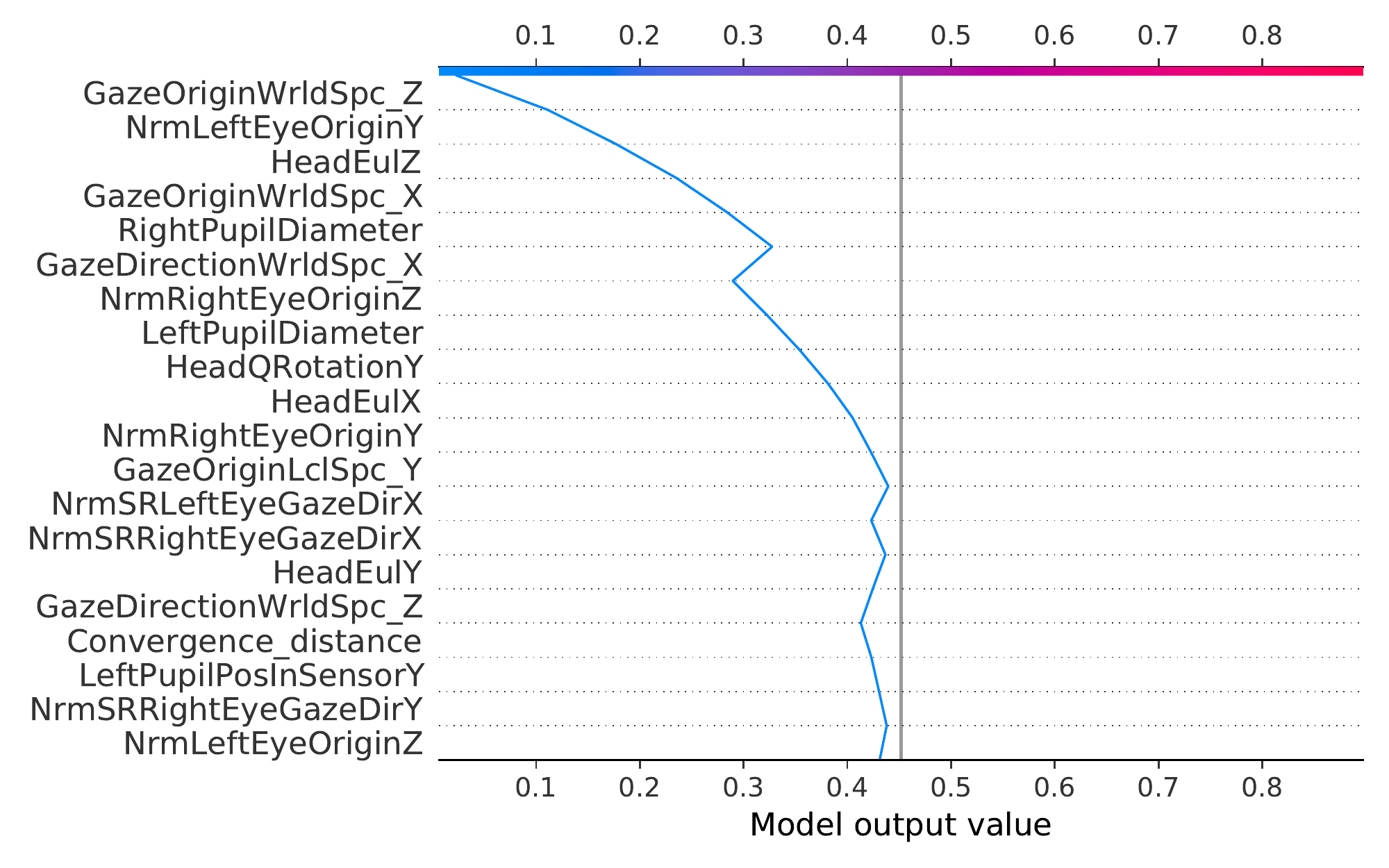}}
\vspace{5pt}
\caption{The local explanation of GRU-based cybersickness classification for (\textbf{a}) explanation for high cybersickness severity, (\textbf{b}) explanation for none cybersickness severity.} \label{fig:local_actual_gru}
\end{figure*}

\begin{figure*}[!t]
\vspace{-0.15in}
\centering      
\subfloat[\centering\label{fig:3a}]{\includegraphics[width=0.5\textwidth]{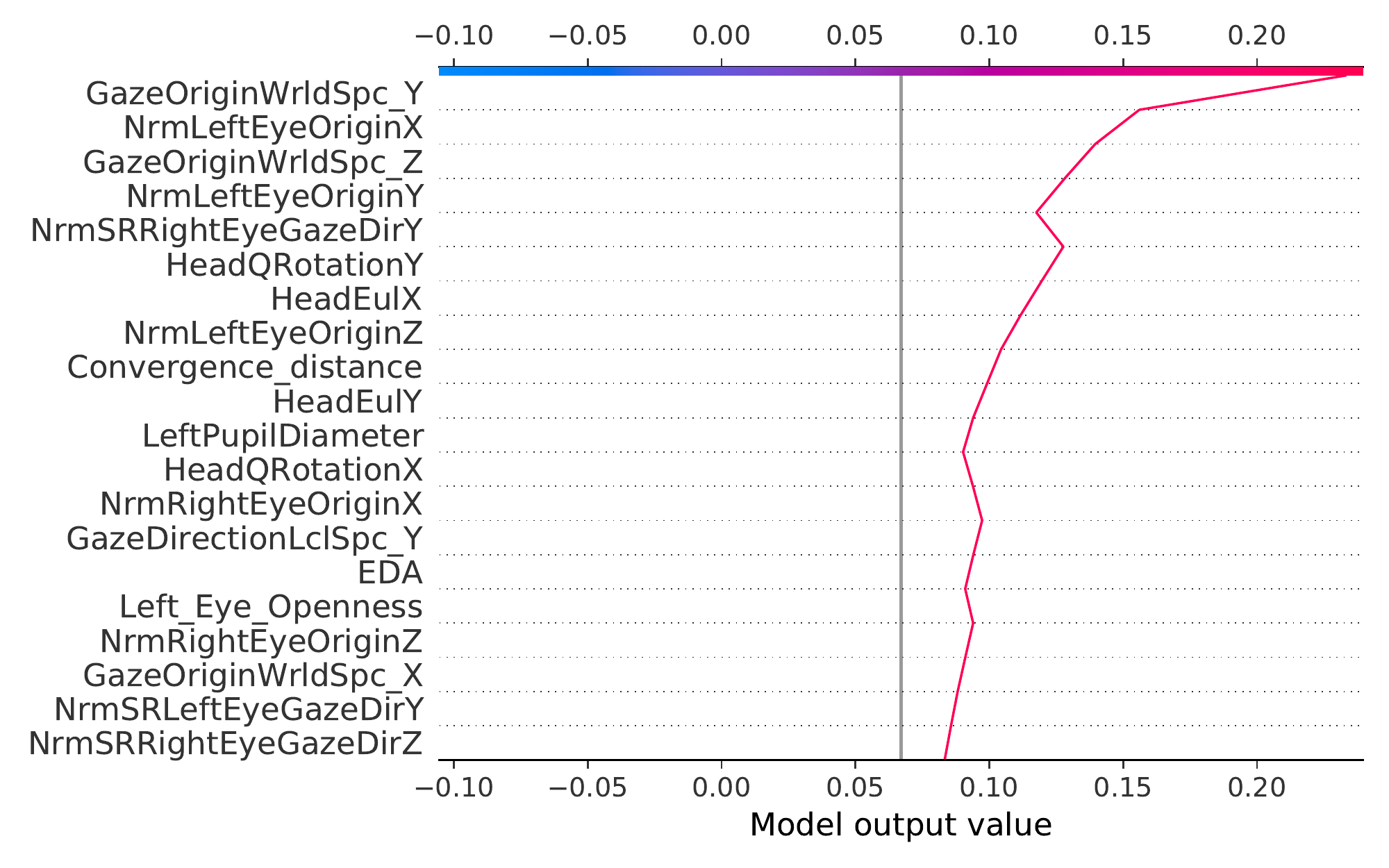}}\hfill
\subfloat[\centering\label{fig:3b}]{\includegraphics[width=0.5\textwidth]{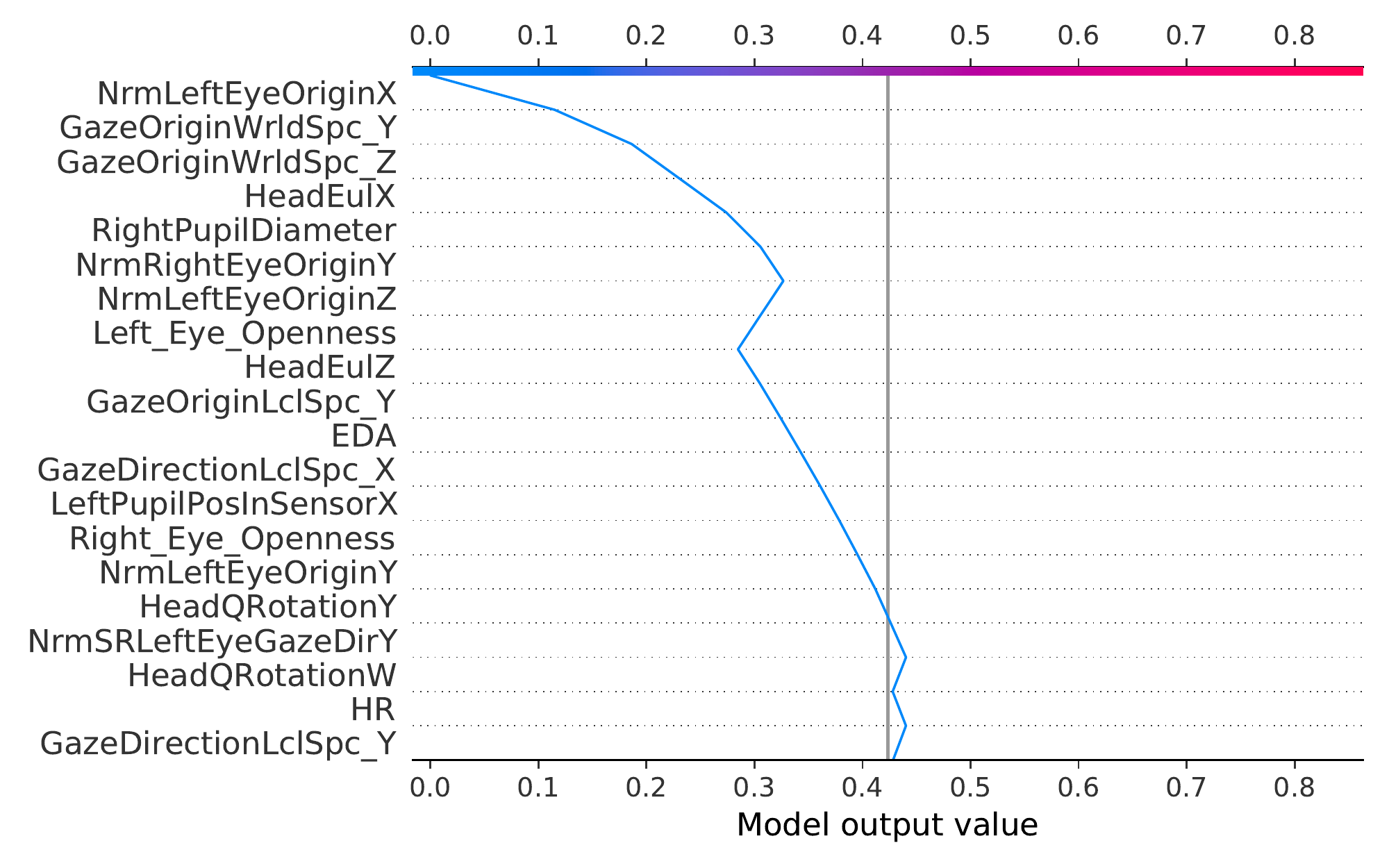}}
\vspace{5pt}
\caption{The local explanation of MLP-based cybersickness classification for the (\textbf{a}) explanation for high cybersickness severity, (\textbf{b}) explanation for low cybersickness severity.} \label{fig:local_actual_mlp}
\end{figure*}

\subsection{Feature space reduction using SHAP-based cybersickness explanation}
In this section, we apply the SHAP-based post-hoc explanation method to explain the DL model's outcome. These explanations, specifically the global explanation, are then used to identify the dominating features to aid in reducing the cybersickness DL models. 

\subsubsection{Cybersickness Severity Global Explanation}
\begin{figure*}[t]
\centering      
\subfloat[\centering\label{fig:3a}]{\includegraphics[width=0.33\textwidth]{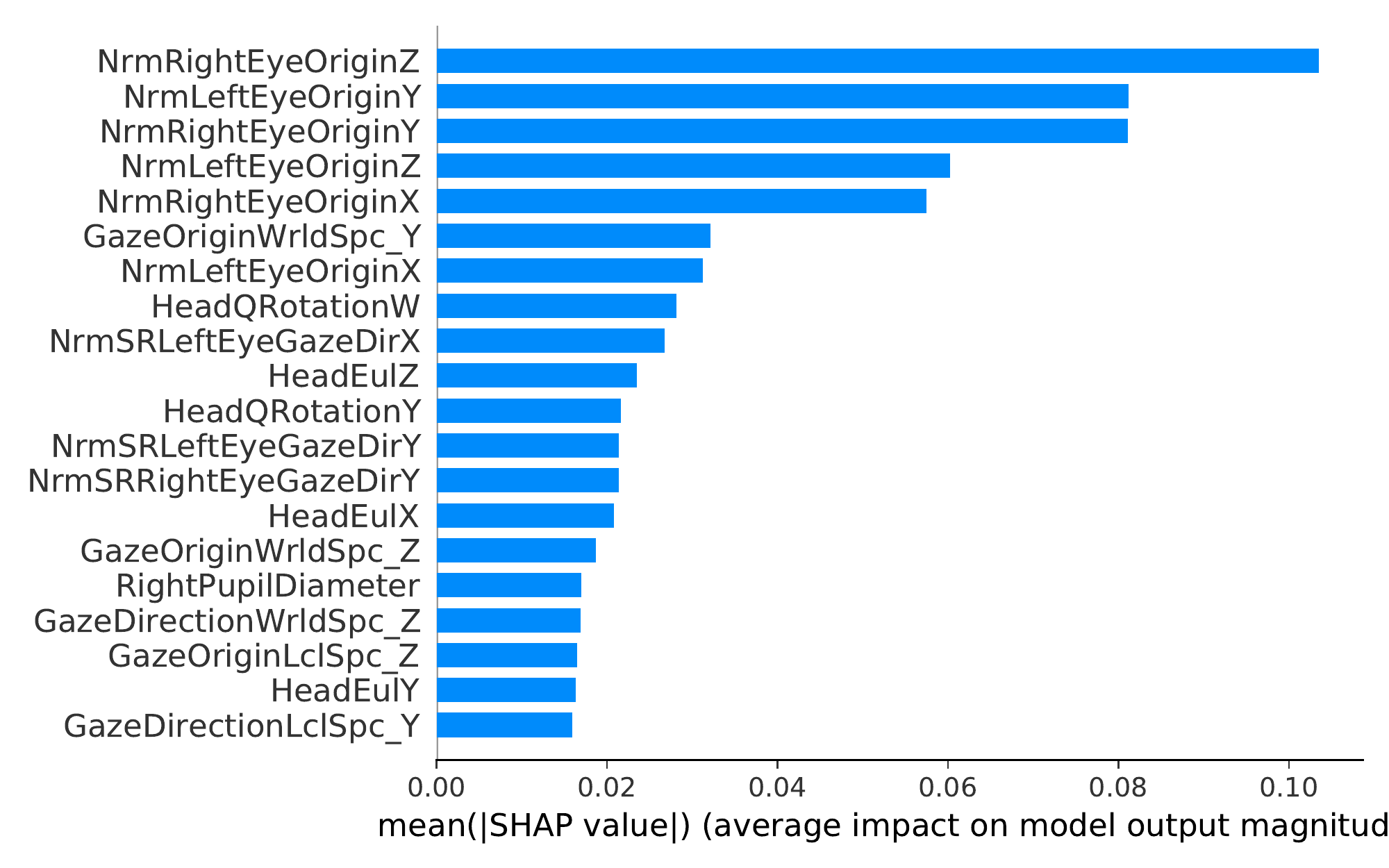}}\hfill
\subfloat[\centering\label{fig:3b}]{\includegraphics[width=0.33\textwidth]{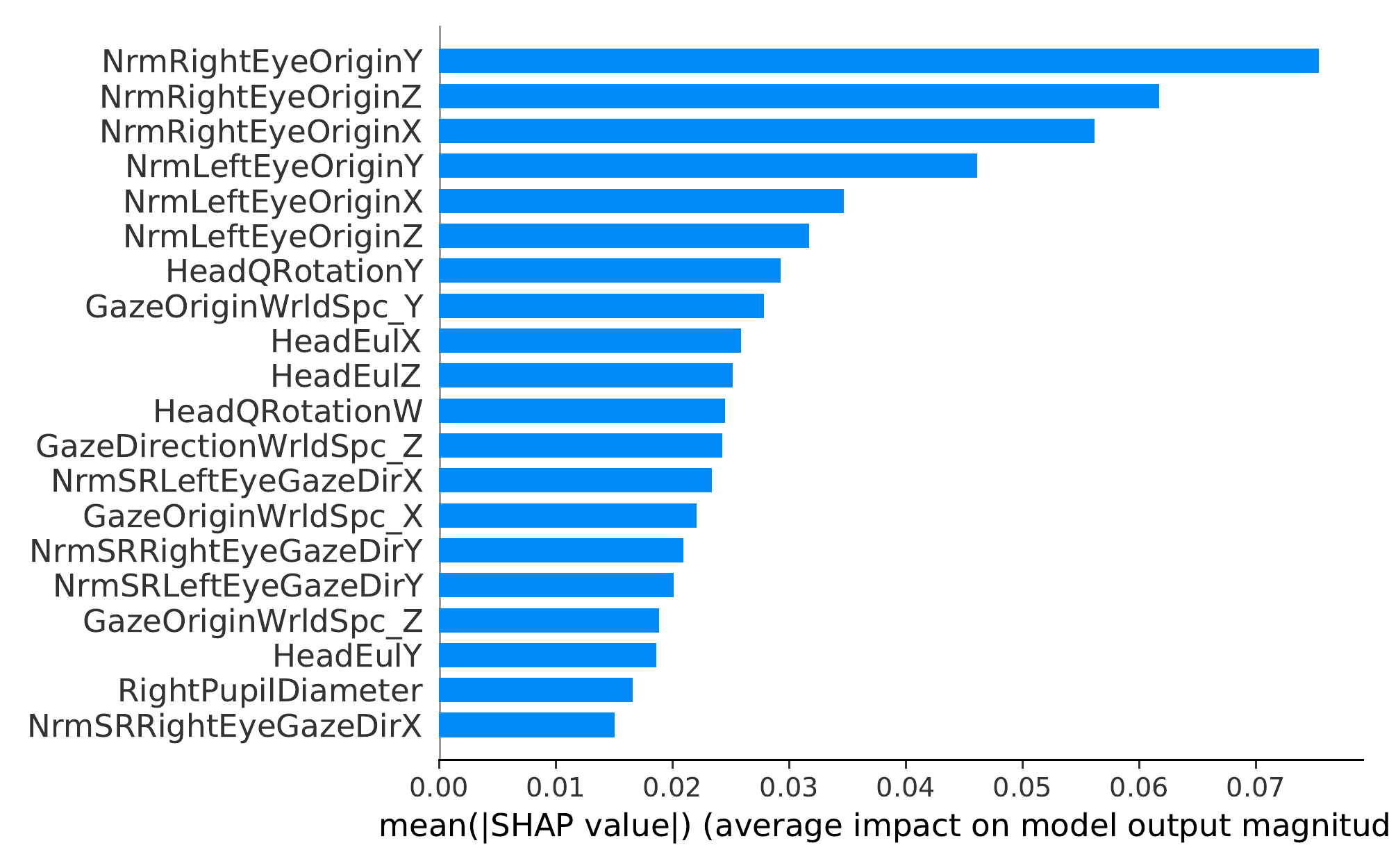}}
\subfloat[\centering\label{fig:3b}]{\includegraphics[width=0.33\textwidth]{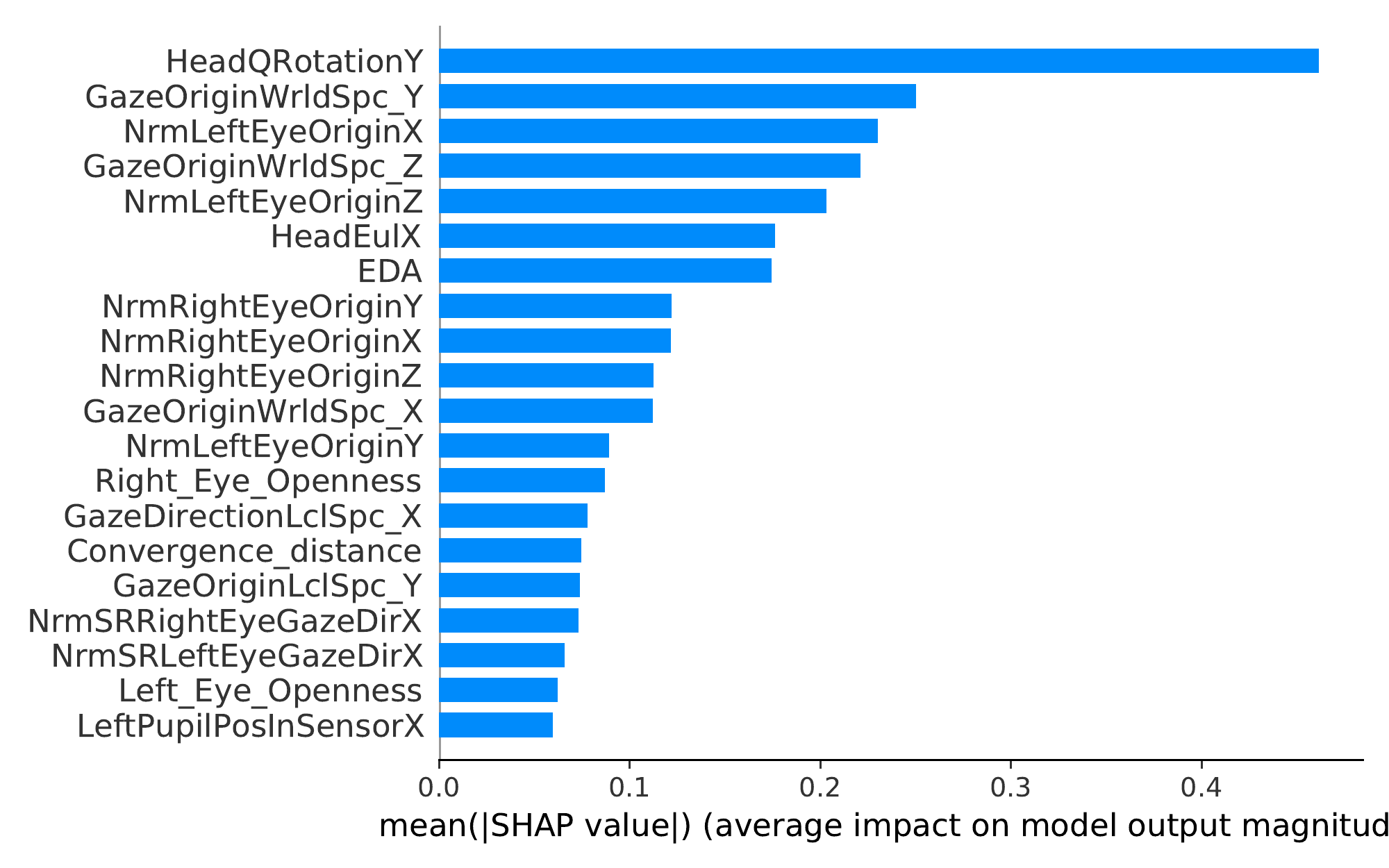}}
\vspace{5pt}
\caption{Overall feature importance using global explanation using SHAP  for (\textbf{a}) LSTM model (\textbf{b}) GRU model (\textbf{c}) MLP model} \label{fig:global_org}
\end{figure*}
The overall feature importance for cybersickness severity classification using LSTM, GRU, and MLP models with all features is visualized in Figure \ref{fig:global_org}. A shapely value is used to calculate the ranking of the most important features contributing to the cybersickness severity classification, with important features at the top and the least important ones at the bottom. The features with large absolute Shapley values are classed as important as they have a higher average impact on the model output. From Figure \ref{fig:global_org}a and b, we observe that features such as \textit{NrmRightEyeOriginZ}, \textit{NrmLeftEyeOriginY}, \textit{NrmRightEyeOriginY}, \textit{NrmLeftEyeOriginZ} , \textit{GazeOriginWrldSpc\_Y}, etc., are the most dominating features in cybersickness severity classification for LSTM and GRU models, which corresponds to the eye tracking features. Interestingly, for both LSTM and GRU models, the eye-tracking features have a much stronger influence on the cybersickness severity classification than head-tracking features. The reason is that eye tracking measurement data contains insightful information such as gaze behavior, the position of the pupil, and the type of blink of the user for tracking the user's activity~\cite{islam2021cybersickness}. Therefore, these eye-tracking features can influence the level of cybersickness. On the other hand, features such as \textit{HeadQRotationY}, \textit{GazeOriginWrldSpc_Y}, \textit{GazeOriginWrldSpc\_Z}, \textit{HeadEulX}, \textit{EDA}, etc., appear to be the most predictive features in cybersickness classification for the MLP model (Figure \ref{fig:global_org}c). The MLP model has many false positives and negatives regarding cybersickness severity classification. However, the cybersickness severity classification using the MLP model has quite a low accuracy compared to LSTM and GRU models, as discussed in Section \ref{subsubsec:class}. Therefore, there is an incorrect ranking of the features such as \textit{HeadQRotationY} at the top of the features list in Figure \ref{fig:global_org}c. Previous work \cite{chang2021predicting, islam2021cybersickness, lopes2020eye, van2022quality} has suggested that eye tracking features are the most influential features in cybersickness severity classification as compared to head tracking features. Such cybersickness explanation about misclassification gives great insights into classification findings and increases trust in the model's prediction.

\subsubsection{Cybersickness Severity Local Explanation}

The individual samples are taken from the model randomly to explain their outcome for cybersickness local explanation. The results of the local explanation utilizing SHAP for the LSTM model are shown in Figure \ref{fig:local_actual_LSTM}. Figure \ref{fig:local_actual_LSTM}a illustrates the high cybersickness severity classification. In that figure, the red-colored line shows cybersickness probabilities for that individual outcome, the $x$-axis represents the model's output as log odds (the probabilities of feature importance in prediction), and the $y$-axis lists the model's features. It is observed that the eye tracking feature \textit{NrmLeftEyeOriginY} is the most influential feature for high cybersickness severity classification, which has the highest probabilities of nearly $1.0$. Most of the features except \textit{HeadQRotationY} have the highest probabilities of outcome belonging to eye tracking features; thus, an appropriate decision is established for high cybersickness severity classification. For example, in Figure \ref{fig:local_actual_LSTM}b, none cybersickness severity classification has low probabilities score for eye tracking features, in which \textit{HeadQRotationW} is the most influential feature with probabilities of $0.4$ and \textit{NrmRightEyeOriginZ} is the least influential feature with probabilities of $0$. As a result, the correct classification is made for none cybersickness severity class. Similarly, the local explanation of the classified cybersickness using the GRU model is shown in Figure \ref{fig:local_actual_gru}. In Figure \ref{fig:local_actual_gru} a and Figure \ref{fig:local_actual_gru} b, we observe that most features contribute to a high cybersickness severity class corresponding to the eye-tracking features and for none cybersickness severity class the dominating features are mostly from head-tracking features i.e., \textit{HeadEulZ}, \textit{HeadEulY}, \textit{CLeftPupilPosInSensorY}, \textit{Convergence_distance}, etc.,. 
Likewise, the local explanation of the classified cybersickness using the MLP model is shown in Figure \ref{fig:local_actual_mlp}. From Figure \ref{fig:local_actual_mlp}a, it is observed that the most influential features for the high cybersickness severity class belong to eye-tracking features such as \textit{GazeOriginWrldSpc\_Y}, \textit{NrmLeftEyeOriginX},etc. However, the features, such as \textit{EDA}, \textit{HeadQRotationY}, \textit{HeadEulX}, etc., have significant importance in high cybersickness severity classification. This observation is misguiding since we know eye-tracking features are more critical in cybersickness prediction. However, this explanation also shows why the accuracy of the cybersickness classification using the MLP model is lower than that of the LSTM and GRU models. Consequently, Figure \ref{fig:local_actual_mlp}b shows that most of the features that influence the positive outcome (none cybersickness severity class) belong to head tracking and physiological signal features again. The \textit{HeadEulZ}, \textit{HeadEulZ}, \textit{HeadQRotationW}, \textit{HR}, etc., are the most predictive features for none cybersickness severity class. Such feature identification provides insights into the classification results and builds trust in the model outcome to make appropriate decisions.

\begin{table*}[ht!]
\centering
\caption{List of selected features for retraining the LSTM, GRU and MLP models through SHAP explanation}
\label{reduced_features}
\resizebox{1.5\columnwidth}{!}{%
\begin{tabular}{|c|c|}
\hline
\textbf{Measurement Type} & \textbf{Selected Features from SHAP Fxplanation} \\ \hline
Eye-Tracking Data & \begin{tabular}[c]{@{}c@{}}NrmRightEyeOriginZ; NrmLeftEyeOriginY; NrmRightEyeOriginX;\\ NrmRightEyeOriginY; NrmLeftEyeOriginZ; NrmLeftEyeOriginX; \\ NrmSRLeftEyeGazeDirX; NrmSRLeftEyeGazeDirY; NrmSRRightEyeGazeDirY;\\ GazeOriginWrldSpc\_Y; GazeOriginWrldSpc\_Z; GazeDirectionWrldSpc\_Z; \\ RightPupilDiameter; GazeOriginLclSpc\_Z\end{tabular} \\ \hline
Head-Tracking Data & HeadQRotationW; HeadQRotationY; HeadEulX; HeadEulZ \\ \hline
\end{tabular}%
}
\end{table*}


\subsection{Model reduction for cybersickness detection}

This section presents the details of feature selection and cybersickness model reduction. \textcolor{blue}{Using the (global) explanation results described in the previous section, we first identify the top 1/3 of the features based on the SHAP explanation in Algorithm \ref{SHAP method}. Once the features are selected we retrain the DL models based on the selected features. We chose 1/3 of the dominating features to develop a reduced model from the base model because we were able to achieve a similar detection accuracy using this ratio. Note for different datasets, this ratio can be different.} It is also worth mentioning that we use the explanation results for GRU and LSTM models as they have excellent accuracies compared to the MLP model. Table \ref{reduced_features} shows the selected features through SHAP explanation for retraining the LSTM, GRU, and MLP models. Table \ref{LSTM_reduced}, Table \ref{GRU_reduced}, and Table \ref{MLP_reduced} show the network architectures of the retrained LSTM, GRU, and MLP models. We observe that the explanation-based model reduction reduces their size significantly. For instance, after reduction, the LSTM model has a total of $19,108$ trainable parameters, which is $6.26$X times less compared to its non-reduced version in \label{LSTM_without}. Similarly, the total number of trainable parameters for the GRU and MLP models are $24,388$ and $4,020$ after model reduction, which is $4$X, and $4.25$X less compared to their non-reduced version in Tables \ref{gru_without} and \ref{mlp_without}, respectively. The reduced models' training and inference times are reported in Table \ref{train_time}. We observe that the training time of the reduced MLP, LSTM, and GRU models are $21.25$, $376.37$, and $774.29$ seconds, which is $2.4$X, $5.6$X, and $4.2$X faster than their non-reduced versions. Similarly, their inference time is also significantly faster after reduction. For instance, the inference time for the reduced MLP, LSTM, and GRU models are $0.09$, $1.07$, $1.32$ seconds, which is $1.66$X, $3.78$X, and $3.72$X faster than their non-reduced versions. In the following subsections, we describe the training and inference performance of the reduced cybersickness DL models.



 
\begin{table}[h]
\centering
\caption{Reduced LSTM model architecture (with reduced features)}
\label{LSTM_reduced}
\resizebox{\columnwidth}{!}{%
\begin{tabular}{|ccc|ccc|}
\hline
\multicolumn{1}{|c|}{\textbf{Layer}} & \multicolumn{1}{c|}{\textbf{Type}} & \textbf{Output shape} & \multicolumn{1}{c|}{\textbf{\#Param}} & \multicolumn{1}{c|}{\textbf{Dropout}} & \textbf{Activation} \\ \hline
\multicolumn{1}{|c|}{1} & \multicolumn{1}{c|}{LSTM} & 64 & \multicolumn{1}{c|}{16896} & \multicolumn{1}{c|}{0.15} & - \\ \hline
\multicolumn{1}{|c|}{2} & \multicolumn{1}{c|}{Dropout} & 64 & \multicolumn{1}{c|}{0} & \multicolumn{1}{c|}{0.15} &  \\ \hline
\multicolumn{1}{|c|}{5} & \multicolumn{1}{c|}{Dense} & 32 & \multicolumn{1}{c|}{2080} & \multicolumn{1}{c|}{-} & ReLU \\ \hline
\multicolumn{1}{|c|}{6} & \multicolumn{1}{c|}{Dense} & 4 & \multicolumn{1}{c|}{132} & \multicolumn{1}{c|}{-} & Softmax \\ \hline
\multicolumn{3}{|c|}{\textbf{Total no. of param:}} & \multicolumn{3}{l|}{\textbf{19,108}} \\ \hline
\end{tabular}%
}
\end{table}


\begin{table}[h]
\centering
\caption{Reduced GRU model architecture (with reduced features)}
\label{GRU_reduced}
\resizebox{\columnwidth}{!}{%
\begin{tabular}{|ccc|ccc|}
\hline
\multicolumn{1}{|c|}{\textbf{Layer}} & \multicolumn{1}{c|}{\textbf{Type}} & \textbf{Output shape} & \multicolumn{1}{c|}{\textbf{\#Param}} & \multicolumn{1}{c|}{\textbf{Dropout}} & \textbf{Activation} \\ \hline
\multicolumn{1}{|c|}{1} & \multicolumn{1}{c|}{GRU} & 32 & \multicolumn{1}{c|}{3360} & \multicolumn{1}{c|}{0.2} & - \\ \hline
\multicolumn{1}{|c|}{2} & \multicolumn{1}{c|}{Dropout} & 32 & \multicolumn{1}{c|}{0} & \multicolumn{1}{c|}{0.2} &  \\ \hline
\multicolumn{1}{|c|}{3} & \multicolumn{1}{c|}{GRU} & 64 & \multicolumn{1}{c|}{18816} & \multicolumn{1}{c|}{0.2} & - \\ \hline
\multicolumn{1}{|c|}{4} & \multicolumn{1}{c|}{Dropout} & 64 & \multicolumn{1}{c|}{0} & \multicolumn{1}{c|}{0.2} & - \\ \hline
\multicolumn{1}{|c|}{5} & \multicolumn{1}{c|}{Dense} & 32 & \multicolumn{1}{c|}{2080} & \multicolumn{1}{c|}{-} & ReLU \\ \hline
\multicolumn{1}{|c|}{6} & \multicolumn{1}{c|}{Dense} & 4 & \multicolumn{1}{c|}{132} & \multicolumn{1}{c|}{-} & Softmax \\ \hline
\multicolumn{3}{|c|}{\textbf{Total no. of param:}} & \multicolumn{3}{l|}{\textbf{24,388}} \\ \hline
\end{tabular}%
}
\end{table}

\begin{table}[h]
\centering
\caption{Reduced MLP model architecture (with reduced features)}
\label{MLP_reduced}
\resizebox{\columnwidth}{!}{%
\begin{tabular}{|ccc|ccc|}
\hline
\multicolumn{1}{|c|}{\textbf{Layer}} & \multicolumn{1}{c|}{\textbf{Type}} & \textbf{Output shape} & \multicolumn{1}{c|}{\textbf{\#Param}} & \multicolumn{1}{c|}{\textbf{Dropout}} & \textbf{Activation} \\ \hline
\multicolumn{1}{|c|}{1} & \multicolumn{1}{c|}{Dense} & 64 & \multicolumn{1}{c|}{1344} & \multicolumn{1}{c|}{0.25} & ReLU \\ \hline
\multicolumn{1}{|c|}{2} & \multicolumn{1}{c|}{Dropout} & 64 & \multicolumn{1}{c|}{0} & \multicolumn{1}{c|}{0.25} &  \\ \hline
\multicolumn{1}{|c|}{3} & \multicolumn{1}{c|}{Dense} & 32 & \multicolumn{1}{c|}{2080} & \multicolumn{1}{c|}{0.25} & ReLU \\ \hline
\multicolumn{1}{|c|}{4} & \multicolumn{1}{c|}{Dropout} & 32 & \multicolumn{1}{c|}{} & \multicolumn{1}{c|}{0.25} & - \\ \hline
\multicolumn{1}{|c|}{5} & \multicolumn{1}{c|}{Dense} & 16 & \multicolumn{1}{c|}{528} & \multicolumn{1}{c|}{-} & ReLU \\ \hline
\multicolumn{1}{|c|}{6} & \multicolumn{1}{c|}{Dense} & 4 & \multicolumn{1}{c|}{68} & \multicolumn{1}{c|}{-} & Softmax \\ \hline
\multicolumn{3}{|c|}{\textbf{Total no. of param:}} & \multicolumn{3}{l|}{\textbf{4,020}} \\ \hline
\end{tabular}%
}
\end{table}

\begin{table*}[]
\caption{Performance of $10$-Fold Cross Validation on Cybersickness Severity Classification using reduced DL models (reduced features)}
\label{reduced_class_table}
\resizebox{\textwidth}{!}{%

\resizebox{\textwidth}{!}{%
\begin{tabular}{|c|cc|c|cccc|cccc|cccc|}
\hline
\multirow{2}{*}{\textbf{Models}} &
  \multicolumn{2}{c|}{\textbf{Feature Count}} &
  \multirow{2}{*} {\textbf{Accuracy\%}} &
  \multicolumn{4}{c|}{\textbf{Precision\%}} &
  \multicolumn{4}{c|}{\textbf{Recall\%}} &
  \multicolumn{4}{c|}{\textbf{F1-score}} \\ \cline{2-3} \cline{5-16} 
 &
  \multicolumn{1}{c|}{\textbf{Original}} &
  \textbf{Reduced} &
   &
  \multicolumn{1}{c|}{\textbf{None}} &
  \multicolumn{1}{c|}{\textbf{Low}} &
  \multicolumn{1}{c|}{\textbf{Medium}} &
  \textbf{High} &
  \multicolumn{1}{c|}{\textbf{None}} &
  \multicolumn{1}{c|}{\textbf{Low}} &
  \multicolumn{1}{c|}{\textbf{Medium}} &
  \textbf{High} &
  \multicolumn{1}{c|}{\textbf{None}} &
  \multicolumn{1}{c|}{\textbf{Low}} &
  \multicolumn{1}{c|}{\textbf{Medium}} &
  \textbf{High} \\ \hline
\textbf{LSTM} &
  \multicolumn{1}{c|}{43} &
  18 &
  94 &
  \multicolumn{1}{c|}{97} &
  \multicolumn{1}{c|}{88} &
  \multicolumn{1}{c|}{90} &
  92 &
  \multicolumn{1}{c|}{94} &
  \multicolumn{1}{c|}{95} &
  \multicolumn{1}{c|}{88} &
  96 &
  \multicolumn{1}{c|}{94} &
  \multicolumn{1}{c|}{93} &
  \multicolumn{1}{c|}{87} &
  91 \\ \hline
\textbf{GRU}&
  \multicolumn{1}{c|}{43} &
  18 &
  93 &
  \multicolumn{1}{c|}{96} &
  \multicolumn{1}{c|}{91} &
  \multicolumn{1}{c|}{85} &
  89 &
  \multicolumn{1}{c|}{96} &
  \multicolumn{1}{c|}{92} &
  \multicolumn{1}{c|}{84} &
  90 &
  \multicolumn{1}{c|}{95} &
  \multicolumn{1}{c|}{93} &
  \multicolumn{1}{c|}{83} &
  88 \\ \hline
\textbf{MLP} &
  \multicolumn{1}{c|}{43} &
  18 &
  89 &
  \multicolumn{1}{c|}{91} &
  \multicolumn{1}{c|}{84} &
  \multicolumn{1}{c|}{82} &
  79 &
  \multicolumn{1}{c|}{93} &
  \multicolumn{1}{c|}{91} &
  \multicolumn{1}{c|}{79} &
  85 &
  \multicolumn{1}{c|}{92} &
  \multicolumn{1}{c|}{88} &
  \multicolumn{1}{c|}{80} &
  84 \\ \hline
\end{tabular}%
}
}
\end{table*}

\subsubsection{Cybersickness classification with reduced DL models}
\label{sec:class_reduced}

Table \ref{reduced_class_table} summarizes the accuracy, precision, recall, and F-1 scores of cybersickness classification using the reduced order LSTM, GRU, and MLP models. For instance, cybersickness classification using the reduced LSTM exhibits $94\%$ accuracy, which is also slightly higher than the accuracy of the non-reduced LSTM model (see Table \ref{acc}). In addition, the cybersickness classification accuracy of the reduced MLP and GRU models also increased by $11.5\%$ and $9.7\%$ compared to their non-reduced versions. Furthermore, other performance metrics, such as precision, recall, and F1-score for the none, low, medium, and high cybersickness classes, slightly increased for the reduced LSTM, GRU, and MLP models compared to their non-reduced versions. For instance, the precision score for the none, low, medium, and high cybersickness classes for the reduced MLP are $97\%$, $88\%$, $90\%$, and $92\%$, which is slightly better than the non-reduced LSTM model. Likewise, the recall score for the none, low, medium, and high cybersickness classes for the reduced MLP model are $93\%$, $91\%$, $79\%$, and $85\%$, which is approximately $1.1$X, $1.06$X, $1.09$X, and $1.15$X higher than the non-reduced MLP model. These results suggest that we significantly improve cybersickness classification with reduced models in all cases. 


\subsubsection{Cybersickness regression with reduced DL models}

Table \ref{reg_rduction} shows the performance of cybersickness regression using the reduced order LSTM, GRU, and MLP models. To illustrate, the MSE, RMSE, MAE, and $(R^2)$ values for the reduced  LSTM and GRU models are $0.18$, $0.30$, $0.17$ and $0.92$ and $0.24$, $0.36$, $0.20$ and $0.88$, respectively. Regarding the reduced MLP, we observe a significant improvement in the MSE, RMSE, $R^2$, and MAE. This is because MLP works more effectively with a small feature set. For example, we observe a decrease of RMSE, MSE, and MAE by approximately $14\%$, $17\%$, and $18\%$, and an increase of $R^2$ by $7.6\%$ for the reduced MLP compared to its non-reduced version. Such improvements are evident in all reduced-order DL models for cybersickness regression.

\begin{table}[]
\centering
\caption{Cybersickness regression using reduced DL models (with reduced features)}
\label{reg_rduction}
\resizebox{\columnwidth}{!}{%
\begin{tabular}{|c|cc|c|c|c|c|}
\hline
\textbf{Regression Models} & \multicolumn{2}{c|}{\textbf{Feature count}} & \textbf{MSE} & \textbf{RMSE} & \textbf{$R^2$} & \textbf{MAE} \\ \hline
\textbf{LSTM} & \multicolumn{1}{c|}{43} & 18 & 0.18 & 0.30 & 0.92 & 0.17 \\ \hline
\textbf{GRU} & \multicolumn{1}{c|}{43} & 18 & 0.24 & 0.36 & 0.88 & 0.20 \\ \hline
\textbf{MLP} & \multicolumn{1}{c|}{43} & 18 & 0.35 & 0.49 & 0.79 & 0.33 \\ \hline
\end{tabular}%
}
\end{table}

\section{Discussion}

The SHAP based global explanation reveals that features such as \textit{normal eye origin}, \textit{gaze origin}, \textit{pupil diameter}, etc., are the most influential features for causing cybersickness. On the contrary, the SHAP-based local explanations of specific predictions offered useful insight for each sample, which helps explain misclassification scenarios. Consequently, we ranked the features' importance from the global explanation using SHAP, and important features were used to retrain DL models. Our results suggest that the SHAP-guided reduced DL models result in significantly faster training and inference time without sacrificing accuracy. For instance, the reduced LSTM cybersickness model, which was trained with only 1/3 of the features compared to its non-reduced version, classified the cybersickness severity with an accuracy of $94\%$. Similarly, while regressing cybersickness, the reduced LSTM obtained an RMSE value of $0.30$, which is $16.8\%$ less than its non-reduced version. The LSTM and GRU models performed great in both cybersickness detection and regression, whereas MLP performed poorly. It is not surprising that the MLP performs poorly in classification and regression since MLP is incapable of remembering the past sequence compared to LSTM and GRU. 
The accuracy of our LSTM models outperforms several state-of-the-art works in DL-based cybersickness detection. For instance, Islam et al.~\cite{islam2021cybersickness} employed the deep fusion model to classify the severity of cybersickness with an accuracy of $87.7\%$ and reported an RMSE value of $0.51$ using eye and head tracking data. The same authors in~\cite{islamtowards} used a deep temporal convolutional network (DeepTCN) to forecast the cybersickness FMS score (on a scale from $0–10$) with an RMSE value of $0.49$, based on eye tracking, heart rate, and galvanic skin response data. Our LSTM model's classification and regression accuracy outperform these works. There also exist other works which are relevant to our work. For instance, Qu et al.~\cite{qu2022bio}, Islam et. al~\cite{islam2020automatic}, Kim et al.~\cite{kim2019deep}, and Jeong et al.~\cite{Jeong2019} reported cybersickness detection accuracy of $96.85\%$, $97.44\%$, $89.16\%$, and $94.02\%$, respectively, using physiological and EEG/ECG signals. 


Even though there is tons of work in cybersickness detection methods, to date, only a few studies have been conducted on identifying the causes of cybersickness~\cite{padmanaban2018towards, kim2019deep, islam2021cybersickness, islam2020deep}. However, to the best of our knowledge, to date the exists no prior work on applying XAI to explain cybersickness DL models and to reduce their dimensions. Indeed, XAI-based explanations can help researchers understand the reasons behind correct and incorrect cybersickness classification and can be further utilized to develop effective cybersickness reduction methods. \textcolor{blue}{On the other hand, using complex DL models for real-time cybersickness detection is computationally intensive and may be impractical for integration into energy-constrained standalone VR HMDs \cite{castaneda2016virtual}. Therefore, we believe that the proposed XAI-based cybersickness model reduction can help deploy those models on standalone HMDs.}

\section{Limitation and future works}

Although the proposed XAI-based DL-based cybersickness detection and reduction, specifically LSTM, outperformed the previous state-of-the-art cybersickness detection models, our approach has a few limitations. \textcolor{blue}{For instance, even though we demonstrated the effectiveness of our proposed XAI-based model reduction method with a fast training and inference time, we did not deploy them on an actual VR headset}. We plan to address this limitation in the future. Furthermore, cybersickness might affect different people based on their unique characteristics, VR environment, and gender \cite{stanney2020virtual}. Therefore, in the future, we plan to conduct further research with people from broader demographic backgrounds and of equal gender representation. Also, this work uses only eye-tracking, head-tracking, and physiological signals to detect cybersickness. In the future, we plan to investigate the effect of stereo images and stereoscopic video data to detect cybersickness with explainability. \textcolor{blue}{Finally, we acknowledge that we do not have access to datasets with more than 7 minutes of exposure time. We plan to develop a new dataset with a long exposure time in the future to validate our approach further}.


\section{Conclusion}
This paper proposed the LiteVR framework, an XAI-based framework for cybersickness detection through lightweight DL models and explanations. Specifically, we developed three cybersickness DL models: MLP, LSTM, and GRU, for cybersickness detection and applied XAI to explain and reduce their size. We illustrated the effectiveness of our proposed method using the integrated sensor dataset with eye-tracking, head-tracking, and physiological signals. Our global explanation results revealed that eye-tracking features are the most influential for causing cybersickness, and using the local explanation, we identified misclassification instances. Furthermore, based on the XAI-based feature ranking, we significantly reduced the model's size (up to 4.3X), training time (up to 5.6X), and inference time (up o 3.8X). For instance, our reduced LSTM cybersickness model could predict cybersickness with an accuracy of  $94\%$ and RMSE of $0.30$, which outperforms the state-of-art works. We believe this research will be helpful for future researchers working on cybersickness detection mitigation and deployment of these models on consumer-level VR headsets. 


\acknowledgments{
This work was funded through grants from the National Science Foundation (CNS 2114035, IIS 2007041, IIS 2211785). Any opinions, findings, conclusions, or recommendations expressed in this publication are those of the authors and do not necessarily reflect the views of the National Science Foundation.}\

\bibliographystyle{abbrv-doi}

\bibliography{template,kah}
\end{document}